\documentclass[%
 aip,
 amsmath,amssymb,
 reprint,%
]{revtex4-1}

\usepackage{graphicx}
\usepackage{csquotes}
\usepackage{hyperref}
\usepackage{dcolumn}
\usepackage{bm}
\usepackage[utf8]{inputenc}
\usepackage[T1]{fontenc}
\usepackage{mathptmx}
\usepackage{etoolbox}
\usepackage[usenames,dvipsnames]{color}

\makeatletter
\def\@email#1#2{%
 \endgroup
 \patchcmd{\titleblock@produce}
  {\frontmatter@RRAPformat}
  {\frontmatter@RRAPformat{\produce@RRAP{*#1\href{mailto:#2}{#2}}}\frontmatter@RRAPformat}
  {}{}
}%
\makeatother
\begin{document}

\preprint{AIP/123-QED}

\title{Stochastic Heat Engine Using a Single Brownian Ellipsoid}

\author{Soham Dutta}

\affiliation{Department of Physics, University Of Calcutta\\92, Acharya Prafulla Chandra Road, Kolkata-700009, India}

\author{Arnab Saha}

\email{sahaarn@gmail.com}
\affiliation{Department of Physics, University Of Calcutta\\92, Acharya Prafulla Chandra Road, Kolkata-700009, India}

\affiliation{Laboratoire de Physique Théorique et Modélisation, UMR 8089, CY Cergy Paris Université, 95302 Cergy-Pontoise, France}

\date{\today}

\begin{abstract}

Optical tweezers can confine position as well as orientation of a Brownian particle by simultaneously exerting restoring force and torque on it. Here we have proposed the theoretical model of a microscopic Stirling engine, using a passive Brownian ellipsoid as its working substance. The position and the orientation degrees of freedom (DoF) of the ellipsoid in two dimensions (2D), both being confined harmonically by the tweezers, are coupled to a hot and a cold thermal bath time-periodically. The stiffness of the force confinement is also time-periodic such that it resembles a piston-like protocol which drives the Brownian ellipsoid through the strokes of a Stirling cycle. The ellipsoid takes heat from the hot bath and partially converts it into useful thermodynamic work. The extracted work and input heat shows explicit dependence on the shape of the working substance as well as its orientational bias. The operational characteristics of the anisotropic Stirling engine is analyzed using the variance in work and efficiency (in the quasi-static regime), where the latter is bounded by both the Carnot limit as well as the isotropic benchmark. Several ways have been proposed to yield maximum efficiency at a minimum fluctuation in the output. The dissipative coupling between the position and orientation of the ellipsoid, that arises due to its spherical-asymmetry (or, shape anisotropy) and a finite mean orientation, plays an important role to optimize the engine characteristics. Finally, we have analytically explored the slightly anisotropic regime, where the coupling is linearized by suitably tuning the system parameters. The average extracted work has also been calculated in this case, which shows an excellent agreement with the numerical results of the fully anisotropic system, when subjected to the stipulated range of parameters.

\end{abstract}

\maketitle

\section{Introduction}

Transportation, manipulation and assembling of colloidal particles are of paramount importance in the realm of microscopic \enquote{machines}. Thermodynamics at such small length-scales (of the order of a micron), which is typically fluctuation-dominated and developed under the framework of stochastic thermodynamics \cite{seifert2012stochastic,sekimoto2010stochastic}, plays a crucial role in the fabrication of stochastic machines --- a field that can also be termed as \emph{micro-engineering}. Geometry of such microscopic objects also plays a pivotal role in their mechanics, which in turn, is important for various aspects of micro-engineering. However, apart from the mechanics, can the geometry of these objects influence its thermodynamics? In this paper, we will address this question with the example of a colloidal micro-heat engine \cite{martinez2017colloidal}, where we will exploit the geometry of its working substance.

Several studies have been carried out in the realm of stochastic thermodynamics \cite{seifert2012stochastic,ciliberto2017experiments}, particularly in the arena of designing stochastic heat engines \cite{martinez2017colloidal} and refrigerators using a single isotropic particle, which can be obtained in the passive as well as active regimes \cite{majumdar2022exactly}. The experimental realization of a microscopic heat engine, consisting of a single, spherically-symmetric colloidal particle confined in a time-dependent optical trap, has been demonstrated in \cite{blickle2012realization,martinez2016brownian}. An extensive analysis of a single-particle stochastic heat engine, constructed by manipulating a Brownian particle in a time-dependent harmonic potential, has been done in \cite{rana2014single}. This study shows that such an engine exhibits qualitative differences in inertial and overdamped regimes. The properties of a microscopic Stirling engine that uses self-propelling particle as a working substance have been studied in \cite{kumari2020stochastic}. The simple model system devised in \cite{saha2019stochastic} can produce thermodynamic work even from active fluctuations, while an engine powered by active dissipation is studied in \cite{saha2018stochastic}. The optimization of an active engine, regardless of the propulsion velocity, is proposed in \cite{gronchi2021optimization}. The simulations in \cite{loos2023nonreciprocal} reveal that it is possible to even construct stochastic refrigerators that can
generate a net heat transfer from a cold to a hot reservoir at the expense of power exerted by the
non-reciprocal forces and thermodynamic \enquote{information}. A Brownian particle can also be driven by a Carnot-type refrigerating protocol, operating between two thermal baths, that reveals anomalous behavior \cite{rana2016anomalous}. The operational characteristics of single-particle heat engines and refrigerators, driven by a time-asymmetric protocol, are studied in \cite{pal2016operational}. The total entropy production and its distribution for a Brownian particle in a harmonic trap, subjected to an external time-dependent force, have been obtained in \cite{saha2009entropy}. This shows that entropy, generally considered as
an ensemble property, can also be applied to a single stochastic trajectory, such that the total entropy production along a single trajectory involves both the particle entropy as well as the entropy change in the environment. The working substance can be driven out of equilibrium by a time-dependent magnetic field as well, the corresponding distribution of the thermodynamic work being given in \cite{saha2008nonequilibrium}. Along with a magnetic field, work can be also be extracted from a single bath beyond the limit set by the second law of thermodynamics, by performing measurement on the system and utilizing the acquired information \cite{pal2014extracting}.

The present work attempts at a major departure from the previous approaches. Our aim is to study how the thermodynamic characteristics of a stochastic heat engine depend on the shape and orientation of its working substance. For this purpose, the suitable candidates that can be employed are the bi-axial particles \cite{han2006brownian,dhont1996introduction} like rods, ellipsoids, sphero-cylinders, etc. The shape of such particles affects their longitudinal and transverse mobilities, along with the axial orientation. This leads to a generic tensorial mobility, the off-diagonal elements of which can be used to couple the translational coordinates of the particle, along with an implicit coupling between its translational and rotational motion. When such a particle (e.g. a Brownian ellipsoid) is trapped in a harmonic confinement (whose stiffness varies periodically with time) set by optical tweezers, and the ambient fluid (in which the colloidal particle is suspended) is alternately connected to two heat reservoirs, the resulting system acts as a stochastic heat engine taking heat from the hot bath and converting it into useful work. Using the tools of stochastic thermodynamics, the energetics of such a heat engine is shown to explicitly depend on the shape parameters and orientational bias of the ellipsoid, which show several qualitative differences from the isotropic version. The orientational bias here is ensured by a restoring torque offered by the tweezers that can \enquote{angularly} trap the particle \cite{la2004optical,deufel2007nanofabricated} .  

The organization of the paper is as follows. First, we have considered a confined Brownian ellipsoid across a plane, subjected to a harmonic trap that offers a force as well as a restoring torque. The spatial coordinates are dissipatively coupled by a finite mobility difference and the finite mean orientation. Then, the major ingredients of the Stirling engine (with the ellipsoid as a working substance) have been discussed. The stiffness of the trap is varied periodically to imitate the isothermal and isochoric steps, such that the bath temperature is switched between two different values. To obtain an analytical benchmark, the stochastic energetics of the isotropic case have been obtained explicitly, using realization-averaged quantities obeying the \enquote{first law} of stochastic thermodynamics \cite{sekimoto2010stochastic,seifert2012stochastic}. The anisotropic system is then studied numerically, and several ways have been proposed to improve the yield of the engine. The thermodynamic quantities show a strong dependence on the shape and orientation of the ellipsoid. Finally, we have obtained the average extracted work in the {\it{slightly}} anisotropic regime, where the dissipative coupling is tuned to be small using physically-motivated approximations. This analytical result agrees well with that of the complete anisotropic system, when the latter is subjected to the stipulated range of parameters (as dictated by the approximations).

\section{Trapped Brownian ellipsoid in 2D --- The Model}

\begin{figure}[htp]
    \centering
    \includegraphics[width=8cm]{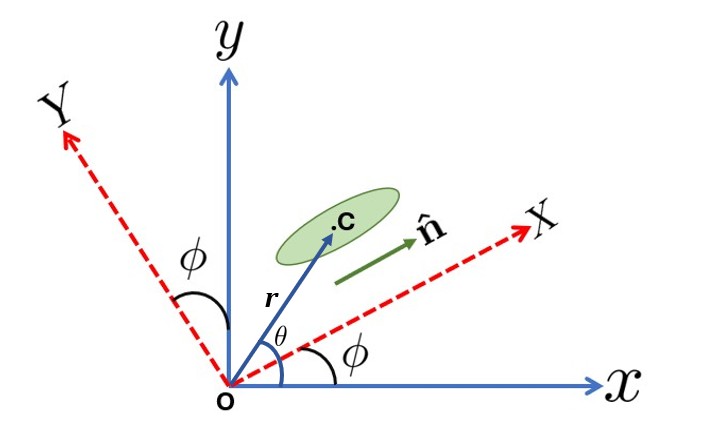}
    \includegraphics[width=8cm]{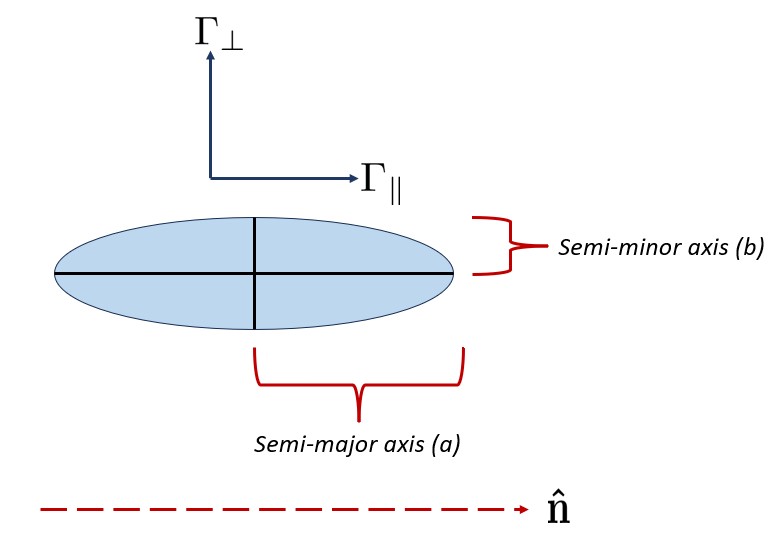}
    \caption{Schematic diagram of the various co-ordinates associated with an ellipsoid. The body frame is denoted by  $(X,Y)$ and the lab frame is denoted by $(x,y)$ --- these frames have been superimposed at a common origin O for visual aid. Here, ${\bf{r}}=\{x,y\}$ is the position vector of the center of mass (C) of the ellipsoid (w.r.t. to the lab frame) and $\theta$ is the corresponding polar angle. The angle $\phi$ denotes the axial orientation of the ellipsoid (w.r.t. the lab frame), the major axis being directed along the unit vector $\hat {\bf n}$. This is also the angle between body and lab axes. The longitudinal ($\Gamma_\parallel$) and transverse ($\Gamma_\perp$) mobilities of the ellipsoid are also depicted respectively along the axial direction and the direction perpendicular to it. For a prolate ellipsoid, the aspect ratio $\alpha=\frac{a}{b}$ is greater than 1.} 
\label{ellipsoid}    
\end{figure}

We consider a colloidal ellipsoid in two dimensions (2D), suspended in a highly viscous fluid where the rate of change of momentum of the particle is negligibly small. Since the dynamics of the particle is restricted to the $xy$-plane (see Fig.[\ref{ellipsoid}]), its center of mass (denoted by C in Fig.[\ref{ellipsoid}]) is moving across this plane only. Furthermore, the particle is free to rotate with the axis being perpendicular to the $xy$-plane and passing through C. This rotation is denoted by the angle $\phi$ in Fig.[\ref{ellipsoid}]. It determines the angular orientation of the particle w.r.t. the lab frame. The overdamped equations of motion of the particle will involve only the coordinates of C at time $t$, denoted by $(x,y)$ and the angle of orientation $\phi$. In Fig.[\ref{ellipsoid}], the coordinates $(x,y,\phi)$, that completely specify the overdamped dynamics of the ellipsoid in 2D, are depicted. The time evolution of all the degrees of freedom (DoF) of the ellipsoid (with respect to the lab-frame) are given by the overdamped Langevin equations as \cite{han2006brownian,ghosh2020persistence,chaki2024dynamics}:

 \begin{eqnarray}
 \nonumber
&&\partial_t x_i=-\sum\limits_{j}\Gamma_{ij}(\phi) \frac{\partial U}{\partial x_j}+\xi_i(t)\\
&&\partial_t \phi=\Gamma_r\tau=\Gamma_r\left[-k_\phi(\phi-\phi_0)+\frac{\xi_r(t)}{\Gamma_r}\right]
\label{eomm}
\end{eqnarray}
where, $i\in \{1,2\}$, $j\in \{1,2\}$, $\phi \in [0:2\pi]$, $x_1\equiv x$, and, $x_2\equiv y$. 

\subsection{Translational dynamics}

According to Eq.[\ref{eomm}], the particle is subjected to three distinct forces during its translation---

(i) The particle is trapped in a spherically-symmetric harmonic potential in 2D, $U(x,y,t)=\frac{k(t)}{2}(x^2+y^2)$, with $k(t)$ being the time-dependent force-constant. This can be thought of as a paraboloidal confinement, whose stiffness changes with time. This will eventually act as a piston-like protocol for the stochastic engine, which will be discussed in the next section. However, this alone cannot couple the translational DoF.

(ii) Next comes the viscous drag  $-\gamma_{ij}v_j$, where $\gamma_{ij}$ is the friction tensor, with $v_j=\partial_tx_j$. In 2D, $\gamma_{ij}$ is a $2\times 2$ matrix, the inverse of which is the mobility tensor represented by $\Gamma_{ij}$. The components of $\Gamma_{ij}$ explicitly depend on the geometry and the orientation angle $\phi$ of the particle as \cite{han2006brownian, kim2013microhydrodynamics, happel2012low, dhont1996introduction,chaki2024dynamics}:

\begin{eqnarray}
\Gamma_{ij}(\phi)&=&\Gamma_{\parallel}(\hat{\bf{n}} \otimes \hat{\bf{n}})+\Gamma_{\perp}[\delta_{ij}-(\hat{\bf{n}} \otimes \hat{\bf{n}})]
\label{mobility}
\end{eqnarray}

\begin{figure}[htp]
    \centering
    \includegraphics[width=10cm]{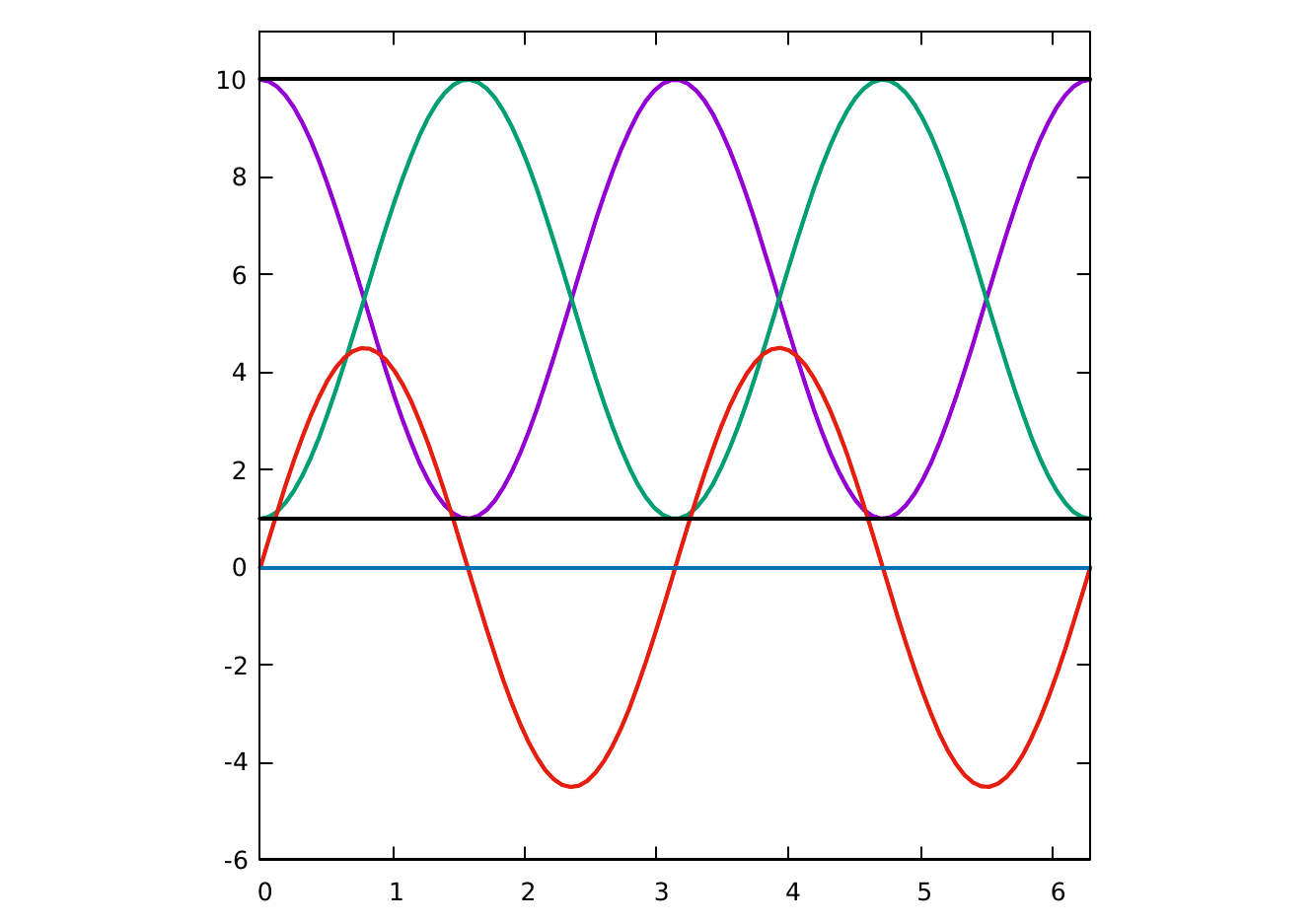}
    \caption{Plot of the elements of mobility tensor versus $\phi\in[0:2\pi]$. The purple, green and red curves denote $\Gamma_{xx}$, $\Gamma_{yy}$ and $\Gamma_{xy}$ respectively. The solid black lines are set at $\Gamma_{\parallel}=10$ and $\Gamma_{\perp}=1$, and the the blue solid line denotes where $\Gamma_{xy}$ becomes zero. Note that when $\Gamma_{xy}$ becomes maximum at $\phi=0.78=\frac{\pi}{4}$, the diagonal elements of the mobility tensor becomes equal, whereas, the contrary happens when $\phi=0$ --- this variation continues periodically with $\phi$. Throughout the paper, this value of $\phi_0$ will be termed as the \emph{maximal coupling regime} (MCR).} 
\label{mobilityvsphi}    
\end{figure}

Here, $\Gamma_{\parallel, \perp}=\frac{1}{\gamma_{\parallel,\perp}}$, where $\gamma_{\parallel}$ and $\gamma_{\perp}$ are the frictional drag coefficients \cite{berg1993random} of the ellipsoidal particle along the directions parallel and perpendicular to its major axis respectively (see Fig.[\ref{ellipsoid}]). Here, $\gamma_{\perp}>\gamma_{\parallel}$, and they depend on the ratio of the lengths of major and minor axes of the ellipsoid \cite{dhont1996introduction,berg1993random}. The mobility difference, $\Delta\Gamma\equiv\Gamma_{\parallel}-\Gamma_{\perp}$, becomes positive as a consequence. The axial vector, $\hat{\bf{n}}\equiv \begin{bmatrix} \cos\phi\\\sin\phi \end{bmatrix}$, is a unit vector (in 2D) along the major axis of the ellipsoidal particle, essentially denoting the local anisotropy in its shape. Explicitly, the mobility tensor then takes the form of a symmetric, $2\times2$ matrix with $\phi$-dependent elements given by (see Fig.[\ref{mobilityvsphi}]):

\begin{eqnarray}
\nonumber
\Gamma_{xx}=\Gamma_{\parallel}\cos^2\phi+\Gamma_{\perp}\sin^2\phi\\
\nonumber
\Gamma_{yy}=\Gamma_{\parallel}\sin^2\phi+\Gamma_{\perp}\cos^2\phi\\
\Gamma_{xy}=\Gamma_{yx}=\Delta\Gamma \sin\phi\cos\phi
\label{gamm}
\end{eqnarray}

One can see that if the particle is spherically symmetric, then $\Delta\Gamma=0$ and $\Gamma_{ij}$ becomes diagonal with $\Gamma_{\perp}=\Gamma_{\parallel}$. Hence, $\Delta\Gamma$ is a measure of the geometric asymmetry of the particle. The Langevin dynamics for $x$ and $y$ will now be coupled with each other via the matrix element $\Gamma_{xy}(\phi,\Gamma_{\parallel},\Gamma_{\perp})$. However, the equation for $\phi$ has no dependence on its translational counterparts, $x$ and $y$. The non-linear (in $\phi$) equations for the translational motion of the Brownian ellipsoid in 2D can now be explicitly given as:

\begin{eqnarray}
\nonumber
\dot{x}&=&-\Gamma_{xx}\frac{\partial U}{\partial x}-\Gamma_{xy}\frac{\partial U}{\partial y}+\xi_x(t)\\
\dot{y}&=&-\Gamma_{yx}\frac{\partial U}{\partial x}-\Gamma_{yy}\frac{\partial U}{\partial y}+\xi_y(t)
\label{eom2}
\end{eqnarray}

(iii) Finally comes the fluctuating forces, $\xi_x(t)$ and $\xi_y(t)$, which are the zero-mean Gaussian, white, random noise components along $x$ and $y$ respectively. The strengths of these noises maintain a fluctuation-dissipation relation (FDR) in the lab-frame\cite{han2006brownian,chaki2024dynamics}, such that (in units of $k_B$):

\begin{eqnarray}
\nonumber
\langle\xi_i (t) \rangle=0\\
\langle\xi_i (t) \xi_j(t')\rangle=2T \Gamma_{ij}(\phi) \delta(t-t')
\label{noise1}
\end{eqnarray}

where, T can be the temperature of the hot or cold bath to which the system will be connected during a complete cycle, as dictated by the piston-like protocol (to be discussed in the next section). Note that the noises along the translational DoF are not independent entities now, but are correlated (in the lab frame) due to $\Gamma_{xy}$. The form of noise-noise correlator given in Eq.[\ref{noise1}] arises due to the transformation of co-ordinates from the body frame of the ellipsoid to the lab frame, such that there exists a relation of the form\cite{chaki2024dynamics,han2006brownian,ghosh2020persistence}:

\begin{eqnarray}
\boldsymbol{\xi}(t)=\sqrt{2}\bar{\bar{D}}(\phi)\boldsymbol{\eta}(t)
\end{eqnarray}

where, $\boldsymbol{\xi}(t)=\{\xi_x(t),\xi_y(t)\}$ and $\boldsymbol{\eta}(t)=\{\eta_X(t),\eta_Y(t)\}$ are the translational noises in the lab and body frames respectively. The quantity $\bar{\bar{D}}(\phi)$ is a $2\times2$, orientation-dependent tensor, given by:

\begin{eqnarray}
\bar{\bar{D}}(\phi)\equiv\sqrt{T}\begin{bmatrix} \sqrt{\Gamma_\parallel}\cos\phi & -\sqrt{\Gamma_\perp}\sin\phi\\\sqrt{\Gamma_\parallel}\sin\phi & \sqrt{\Gamma_\perp}\cos\phi \end{bmatrix}   
\end{eqnarray}

which can also be defined as the so-called \enquote{diffusivity} tensor, as dictated by Einstein's relation. This ensures that the translational DoF are equilibrated to the bath temperature at large times. The noises, $\eta(t)$, are the ones experienced by the ellipsoid while translating in its body frame, which are simply Gaussian, white, uncorrelated, random noises with zero mean:

\begin{eqnarray}
\nonumber
\langle\eta_p (t) \rangle&=&0\\
\langle\eta_p (t) \eta_q(t')\rangle&=&\delta_{pq} \delta(t-t')
\label{noise2}
\end{eqnarray}

with, $p\in\{X,Y\}$ and $q\in\{X,Y\}$.

\subsection{Orientational dynamics}

The dynamics of the axial orientation $(\phi)$ of the ellipsoid as described in Eq.[\ref{eomm}] is governed by a net torque, $\tau$, which is the sum of two torques --- a deterministic one and a thermally-generated, stochastic one. The former is a restoring torque, which always tries to bring the orientation of the particle to a non-zero mean value $\phi_0$, with $k_\phi$ being the constant torque-strength. This torque can be taken as a conservative one, derivable from a $\phi$-dependent confinement of the form, $U_r(\phi)=\frac{k_\phi}{2}(\phi-\phi_0)^2$, that can solely be applied to the body frame of the ellipsoid (the angular orientation has no counterpart in the lab frame). It must be noted that, as the ellipsoid is taken to be passive, this torque is applied externally to the ellipsoid in its body frame. This torque can also have a chiral origin, which is beyond the scope of this paper.

The latter is a random torque, which is simply a Gaussian, white, uncorrelated random noise with zero mean:

\begin{eqnarray}
\nonumber
\langle\xi_r (t) \rangle=0\\
\langle\xi_r (t) \xi_r(t')\rangle=2\Gamma_r T \delta(t-t')
\label{noise}
\end{eqnarray}

where, $\Gamma_r=\frac{1}{\gamma_r}$ is the (scalar) mobility coefficient. The rotational mobility is introduced due to the viscosity of the surrounding fluid in which the particle allows itself to rotate. As the dynamics of $\phi$ is governed by a linear equation, the steady-state distribution of this random variable will also be Gaussian, that is, the first two moments of $\phi$ --- namely, mean $(\langle\phi\rangle)$ and variance $(\sigma_\phi^2)$, are sufficient to quantify its equilibrium distribution. At large times, these moments can be easily evaluated as:

\begin{eqnarray}
\nonumber
\langle\phi\rangle=\phi_0\\
\sigma_\phi^2 \equiv \langle\phi^2\rangle-\langle\phi\rangle^2 = \frac{T}{k_\phi}
\label{phimom}
\end{eqnarray}

which evidently shows that a large value of the restoring torque-strength suppresses the fluctuations in $\phi$, for a given value of the bath temperature $(T)$. For a non-zero value of $\phi_0$, the Gaussian distribution becomes asymmetric about $\phi=0$. This, along with $\Delta\Gamma\neq0$, will drive the system away from the isotropic regime --- as manifested by the corresponding stochastic energetics (to be discussed later).

The application of the deterministic torque is an experimental nuance in itself. Typically, in the colloidal scale, forces and torques are applied on a particle by irradiating it with electromagnetic waves \cite{simpson2007optical,ling2010optical,simpson2011computational,callegari2015computational,loudet2014optically,barton1989theoretical,borghese2008radiation,roy2016using}, via a laser trap --- the arrangement is known as the so-called \emph{optical tweezer}. The force generated from radiation pressure is used to counteract the weight of the particle, thus localizing it to a desired point. By varying the intensity of the beam, a force-gradient can also be generated, such that the particle is always \enquote{pushed} towards the beam focus (that is, where the intensity attains a maximum value). This restoring force creates a spatial confinement for the particle. For anisotropically-shaped particles (e.g. rod, ellipsoid, sphero-cylinder, etc.), this force can even generate a torque to restore the orientation of the particle to a desired value which is known as angular trapping \cite{la2004optical,deufel2007nanofabricated,simpson2007optical,simpson2011computational}. This is also an indicator of the inherent coupling of translational and rotational motions, which occurs only for particles with anisotropic shape \cite{roy2016using,han2006brownian}. It has also been shown that this torque-strength has an explicit relation with the peak electric field in the radiation along with the length-scales of the particle under consideration.

\section{Stochastic Heat Engine --- The Time-dependent Protocol }

Heat engines convert heat into useful work, obeying the thermodynamic propositions. They are connected cyclically between two heat baths, kept at
different temperatures --- the generic scheme of operation of such engines is depicted in Fig.[\ref{enginescheme}]. The second law of thermodynamics bounds their efficiency to the well-known Carnot limit. However, this efficiency can only be achieved in the quasi-static limit, where the transitions between thermodynamic states (availed by the system) occur infinitesimally slowly and hence the output power vanishes. Contrary to this, an efficient engine at maximum power output is governed by the limit set by Curzon and Ahlborn \cite{curzon1975efficiency}.

\begin{figure}[htp]
    \centering
    \includegraphics[width=8cm]{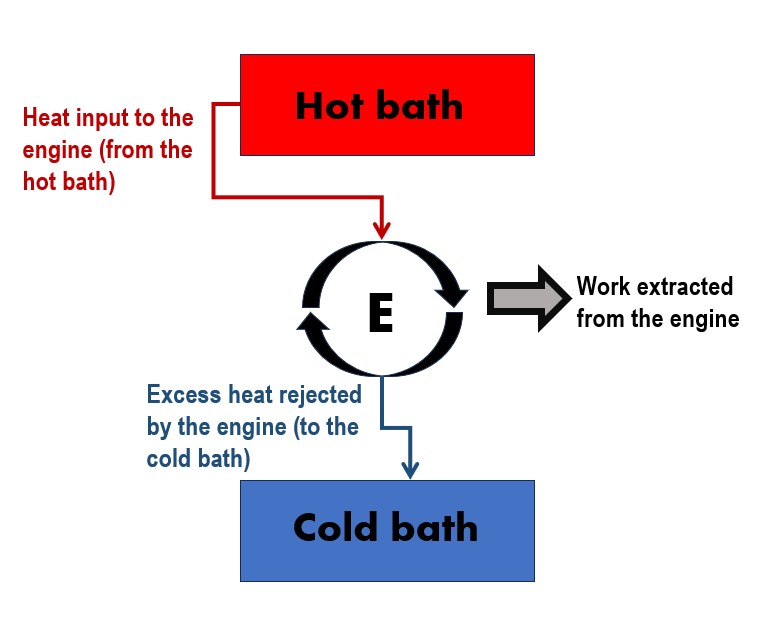}
    \caption{General scheme of operation of a thermodynamic engine (E), cyclically operating between a hot and a cold bath, to enable the extraction of work.} 
\label{enginescheme}    
\end{figure}

In our case, a single ellipsoid is trapped in a 2D, harmonic trap, whose stiffness constant varies periodically with time, in accordance with a protocol. This confinement can be thought of as a \enquote{breathing paraboloid}, that controls the extent of the planar region that can be availed by the particle (or, the \enquote{volume} traversed by the ellipsoid during its translation). In the Brownian regime, this protocol imitates the action of a piston in the macroscopic counterpart. The ambient fluid in which the colloidal ellipsoid resides acts as the thermal reservoir of this engine, whose temperature can be switched between two distinct values (as synchronized with the protocol), thus imitating the presence of a hot and a cold bath. The experimental nuances associated with the practical realization of a stochastic heat engine, namely particle-trapping, heating and temperature control, were conducted using precise laser arrangements by Bechinger et. al. \cite{blickle2012realization}

A Stirling engine contains four steps in a complete cycle of operation --- two isothermal and two isochoric. A quasi-static change in the stiffness of the confining trap can mimic the isothermal steps, whereas, the isochoric (constant-volume) arms can be modeled by an sudden change in the temperature of the bath, without any change in stiffness constant. Our present study will be based on this model \cite{kumari2020stochastic}. 

\begin{figure}[htp]
    \centering
    \includegraphics[width=9cm]{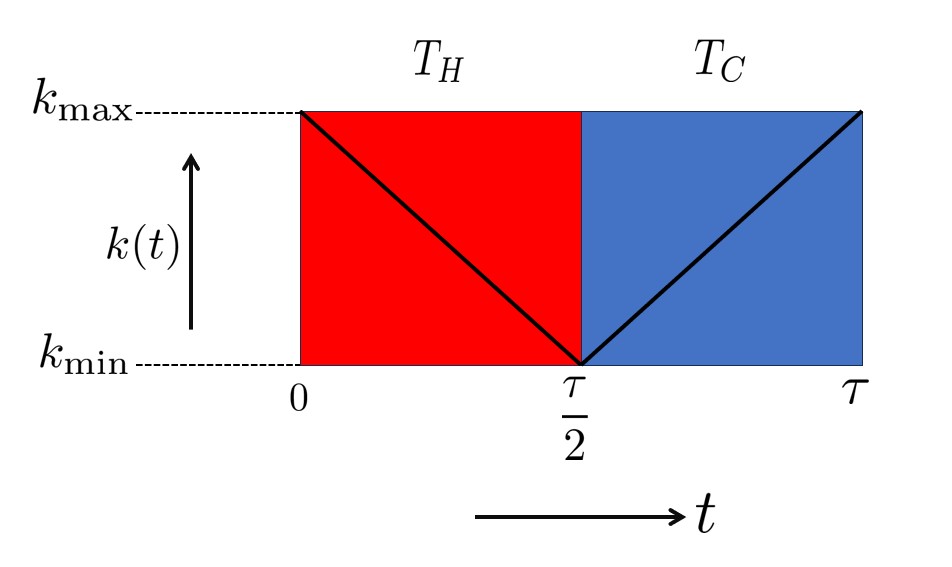}
    \caption{Schematic diagram of the time-dependence of stiffness constant, $k(t)$.} 
\label{protocol}    
\end{figure}

The cycle is switched on at $t=0$ and completed at $t=\tau$. In the isothermal expansion process $\left(0\leq t \leq \frac{\tau}{2}\right)$, the bath temperature is set to $T=T_H$ (hot bath) and the stiffness is linearly decreased (with time) from $k_\text{max}$ to $k_\text{min}$ (see Fig.[\ref{protocol}]), as per the relation \cite{saha2019stochastic}:

\begin{eqnarray}
k(t)=k_\text{max}+(k_\text{min} - k_\text{max})\frac{2t}{\tau}
\end{eqnarray}

In the second step, the temperature is suddenly decreased to $T=T_C$ (cold bath), with the value of stiffness constant held fixed at $k_\text{min}$. This isochoric jump occurs at $t=\frac{\tau}{2}$.

The third step is that of an isothermal compression in contact with the cold bath $\left(\frac{\tau}{2}< t \leq \tau\right)$, where the stiffness constant is linearly increased to $k_\text{max}$ again (see Fig.[\ref{protocol}]), as per the following relation \cite{saha2019stochastic}:

\begin{eqnarray}
k(t)=k_\text{min}+\left(k_\text{min} - k_\text{max}\right)\left(1-\frac{2t}{\tau}\right)
\end{eqnarray}

The cycle concludes with the fourth step, which is the second isochoric jump at $t=\tau$, in which the temperature is suddenly increased to $T=T_H$ again, with the value of stiffness constant held fixed at $k_\text{max}$. This completes the Stirling cycle, which can now be executed periodically and reversibly to obtain useful thermodynamic work. For the quasi-static limit, the cycle time $\tau$ is taken to be large, along with a small value of the constant time-rate of change of $k(t)$ --- this dictates that $(k_\text{max}-k_\text{min})$ must not be very large. The schematic representation of all the steps has been depicted in Fig.[\ref{enginescheme}].

\begin{figure}[htp]
    \centering
    \includegraphics[width=10cm]{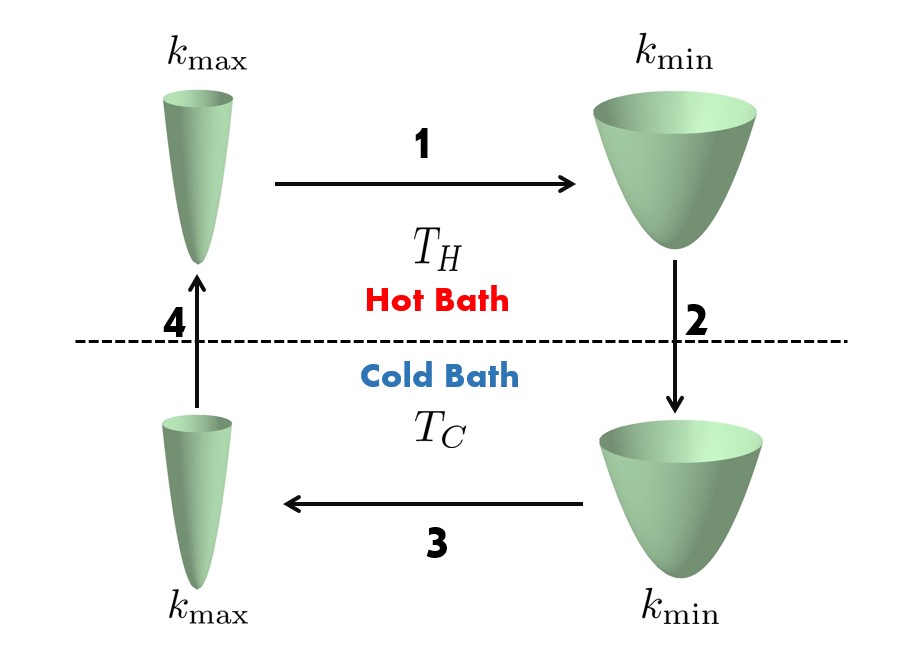}
    \caption{Schematic diagram of the steps in the Stirling cycle implemented in our model : 1 --- isothermal expansion, 2 --- first isochoric jump, 3 --- isothermal compression, 4 --- second isochoric jump.} 
\label{enginescheme}    
\end{figure}

As discussed earlier, the force and torque are simultaneously exerted on the particle via the radiation pressure originating due to electromagnetic waves. The protocol, discussed above, can be realized by tuning the intensity of the radiation, that is, the peak electric field can be set to be time-dependent. Irrespective of shape, these external drives can be exerted to spherical as well as non-spherical particles, with the latter only being vulnerable to the restoring torque that tries to align its orientation to a non-zero, mean value (as required in our case). Nonetheless, the radiation force and torque can be applied simultaneously to spherical particles as well, a detailed study of which has been done in the seminal work \cite{barton1989theoretical}. Here, it has been explicitly shown that the ratio of the average radiation torque and force is proportional to the radius of the spherical particle experiencing these drives. Needless to say, this length scale will be replaced by the lengths of the major and minor axes, in case of an ellipsoid. For a Brownian ellipsoid, these length scales will be of the order of a micron. Besides this, the time-variation of force is infinitesimally slow in the quasi-static limit. As a consequence, the time-variation of the torque will occur much slower than that of the force. Owing to these physical premises, the torque strength $k_\phi$ has been assumed to be a constant (over time) throughout our analysis, even when a time-dependent protocol has been assigned to the stiffness constant, $k(t)$ --- though both of them have the same electromagnetic origin.

\section{Stochastic Thermodynamics and Isotropic limit}

\subsection{Work}

As the system is driven by a time-dependent protocol,  for a given realization of the cyclic process facilitated by the protocol, the evolution of the system is governed by the stochastic trajectory $\{x(t),y(t),\phi(t)\}$, of the ellipsoid in the lab frame, and the work extracted from the system is the time integral of the explicit time-rate of change of the trapping potential, $U(x,y,t)$, along this trajectory \cite{jarzynski1997nonequilibrium,seifert2012stochastic}:

\begin{eqnarray}
    W\equiv\int \frac{\partial U(x,y,t)}{\partial t}dt
    \label{jar}
\end{eqnarray}
Now one can imagine an ensemble of several equivalent realizations of the complete cycle, with initial conditions being generated from a canonical ensemble (here, for a given $k$ and $T$). Then, the average of the stochastic work $W$ may be computed as an average over many such realizations, denoted by $\langle\dots\rangle$ as:

\begin{eqnarray}
    \langle W \rangle = \frac{1}{2}\int \dot{k}[\langle x^2 \rangle + \langle y^2 \rangle] dt
    \label{workdef}
\end{eqnarray}

where, the position moments have been evaluated for the steady-state and an overhead dot denotes a single derivative w.r.t. time. Eq.[\ref{workdef}] is a generic prescription that can be used to evaluate work for both isotropic as well as anisotropic cases. The latter is analytically non-trivial due to the number of involved system parameters and multiplicative noises --- instead, we will address the anisotropic case numerically. 

In the isotropic (spherical) regime, $\Delta\Gamma$ becomes zero (that is, $\Gamma_{xx}=\Gamma_{yy}=\Gamma$). This eradicates the dissipative coupling between the translation DoF os the system, and Eq.[\ref{eom2}] becomes:

\begin{eqnarray}
\nonumber
\dot{x}&=&-k(t)\Gamma x + \sqrt{2\Gamma T}\eta_x(t)\\
\dot{y}&=&-k(t)\Gamma y + \sqrt{2\Gamma T}\eta_y(t)
\label{eom3}
\end{eqnarray}

As $k_{\phi}$ is time-independent,  the orientational dynamics becomes redundant for a passive sphere in the context of computing $W$. Eq.[\ref{eom3}] contains a set of linear, decoupled differential equations, and the statistical properties of the fluctuations are well-defined (see Eq.[\ref{noise2}]). Using the formal solution for the stochastic trajectory, $\{x,y\}$, the position moments can be easily obtained (for quasi-static case) as:

\begin{eqnarray}
\nonumber
\langle x \rangle = 0 = \langle y \rangle\\
\langle x^2 \rangle \equiv \sigma_x = \frac{T}{k(t)} = \langle y^2 \rangle \equiv \sigma_y
\label{moments}
\end{eqnarray}

The work extracted from the stochastic engine is due to the contributions coming from both the isothermal steps --- a decrease in trap stiffness when connected to the hot bath, and an increase in the same when connected to the cold bath. Keeping these thermodynamic processes in mind, the total isothermal contribution to the average extracted work (in the isotropic limit, during an entire cycle) can be calculated (in units of $k_B$) by plugging Eq.[\ref{moments}] in Eq.[\ref{workdef}], to obtain:

\begin{eqnarray}
    \langle W \rangle = (T_C - T_H)\text{ln}\left(\frac{k_\text{max}}{k_\text{min}}\right)
    \label{isowork}
\end{eqnarray}

Owing to the commonly-used sign convention, this expression clearly indicates that work can be extracted out of the isotropic system (a spherical particle in a time-dependent trap) as well, the difference in bath temperatures and the protocol being pivotal ingredients. Also, it can be noted that the explicit time-variation of the trap becomes redundant in the calculation of work and only the extreme values of $k(t)$ were required --- this is a direct consequence of the assumed quasi-static nature of driving protocol.

\subsection{Heat and \enquote{first law}}

Eq.[\ref{eomm}] can be seen as a force/torque-balance equation, where the (internal) bath-generated forces and torques are compensated solely by the (external) trapping force/restoring torque, such that the system remains in mechanical equilibrium besides thermal equilibrium. This becomes clearer if we rewrite the equations in the following form:

\begin{eqnarray}
\nonumber
    \sum\limits_{j}\gamma_{ij}(-\partial_t x_j + \xi_j)=k(t) x_i\\
    -\gamma_r \dot{\phi}+\xi_r(t)=k_\phi(\phi-\phi_0)
\end{eqnarray}

where, the left-hand side bears all the forces and torques originating from the bath itself. The heat associated with the system can now be simply obtained as the time-integral of the power associated with these drives associated with the bath. The total heat $(Q)$ is made up of contributions coming from both translational $(Q_T)$ as well as orientational $(Q_R)$ dynamics, such that $Q=Q_T + Q_R$, where:

\begin{eqnarray}
\nonumber
    Q_T\equiv\int k(t) [x\dot{x}+y\dot{y}] dt\\
    Q_R\equiv k_\phi \int \phi \dot{\phi} dt
\end{eqnarray}

As discussed earlier, an average over several realizations of the system can also be taken for the stochastic quantity $Q$ to obtain a meaningful thermodynamic description. On taking this average and evaluating the above integrals, the total average heat $(\langle Q \rangle)$ can be obtained as the difference between the total change in average internal energy $(\Delta\langle E \rangle)$ of the system (which is just the spatial and angular potential energy) and the total average work $(\langle W \rangle)$, that is:

\begin{eqnarray}
     \langle Q \rangle =\underbrace{\Delta\left[\frac{k(t)}{2}\{\sigma_x + \sigma_y\} +\frac{k_\phi}{2}\langle\phi^2\rangle\right]}_{\Delta\langle E \rangle} - \langle W \rangle
     \label{firstlaw}
\end{eqnarray}

where, the definition of $\langle W \rangle$ is given in Eq.[\ref{workdef}]. The above expression is the well-known \enquote{first law} of stochastic thermodynamics, as prescribed by Sekimoto in his monograph \cite{sekimoto2010stochastic}. Needless to say, stochastic thermodynamics systematically provides a paradigm for carefully extending the notions of classical thermodynamics such as work and heat to the level of individual trajectories of non-equilibrium ensembles. It applies whenever a non-equilibrium system is coupled to one (or several) heat bath(s) of constant temperature. These exact results become particularly relevant for small systems with appreciable thermal fluctuations. For such systems, a first law-like energy balance can be identified along stochastic trajectories.

Eq.[\ref{firstlaw}] can be generically used to evaluate heat for both isotropic as well as anisotropic cases. We will numerically analyze the anisotropic case and the isotropic case will be studied analytically. For the isotropic case $(\Delta\Gamma=0)$, no orientation-restoring torque acts on the passive spherical particle, that is $k_\phi=0$. Hence, while calculating heat in the isotropic case we need to consider both $\Delta \Gamma$ and $k_{\phi}$ to be zero. Now, heat is absorbed by the system during the first step of isothermal expansion --- while the system remains connected to the hot bath, the trap \enquote{expands} due to a linear decrease in the stiffness constant and a heat flux is subjected to the colloidal ellipsoid (acting as a working substance) from the ambient fluid. This \enquote{input} heat $(\langle Q_1\rangle)$ that goes into the system (undergoing a Stirling cycle) is made up of two distinct constituents : one comes from the work done by the ellipsoid during the first isothermal step $(\langle W_1 \rangle)$ and the other comes due to the change in the internal energy of the system $(\Delta\langle E_1\rangle)$ after the second isochoric jump in the bath temperature. In the isotropic limit, the first contribution can be calculated using Eq.[\ref{workdef}] as:

\begin{eqnarray}
    \langle W_1 \rangle = T_H \text{ln} \left(\frac{k_\text{min}}{k_\text{max}}\right)
\end{eqnarray}

The second contribution requires some discussion. At $t=0^-$, the system was in contact with the cold bath $(T_C)$, whereas, at $t=0^+$, the system comes in contact with the hot bath $(T_H)$ --- thus, the system has to relax into a new equilibrium, after a sudden change in temperature (see Fig.[\ref{protocol},\ref{enginescheme}]). The time taken for this relaxation process is assumed to be negligible, as compared to the cycle time $(\tau)$. This relaxation causes an additional heat flow, leading to a change in the internal energy of the system \cite{rana2014single}. The definition of the change in internal energy can be borrowed from Eq.[\ref{firstlaw}], where the change will only be considered along the first isotherm as per the requirements of calculating the input heat. Using the variances in position given by Eq.[\ref{moments}], this change in internal energy can be readily evaluated as:

\begin{eqnarray}
    \Delta\langle E_1 \rangle = T_H - T_C
\end{eqnarray}

Hence, using the first law-like prescription discussed earlier, the input heat (that goes into the system during isothermal expansion) for the isotropic system comes out to be:

\begin{eqnarray}
    \langle Q_1 \rangle = (T_H - T_C)+T_H \text{ln} \left(\frac{k_\text{max}}{k_\text{min}}\right)
    \label{isoheat}
\end{eqnarray}

which will always be positive according to the sign convention.


\subsection{Quasi-static efficiency}

As dictated by the second law of thermodynamics, not all of the input heat can be recovered and extracted as useful work. Hence, efficiency of a stochastic heat engine becomes an important quantifier to quantify its performance. In our case, efficiency of the Stirling engine (in the quasi-static limit) can be computed as the ratio of the absolute value of the average extracted work (obtained from the complete cycle) to the average input heat (during the first isotherm), which can be obtained as:

\begin{eqnarray}
    \eta_\text{Stirling}\equiv\frac{|\langle W \rangle|}{\langle Q_1 \rangle} = \frac{\eta_C \text{ln}(\kappa)}{\eta_C + \text{ln}(\kappa)}
    \label{isoeffi}
\end{eqnarray}

where, $\kappa\equiv\frac{k_\text{max}}{k_\text{min}}$ and $\eta_C = 1-\frac{T_C}{T_H}$ is the well-known Carnot efficiency. Clearly, this definition of efficiency is not that of a fluctuating quantity --- nevertheless, efficiency can be stochastic, in general. As evident from Eq.[\ref{isoeffi}], $\eta_\text{Stirling}< \eta_C$ for finite values of $\kappa$ --- this was expected as no heat engine, working between two heat baths, can be more efficient than a reversible, Carnot engine working between the same baths.

Having acquired the essential tools of stochastic thermodynamics along with the results for the complete isotropic system, we can now proceed towards the numerical analysis of the anisotropic system and fetch the set of system parameters where the isotropic benchmarks can be reproduced.

\section{Results --- Fully Anisotropic case}

\subsection{General scheme of simulation and isotropic benchmarks}

The Langevin equations for translation and orientation, as given in Eq.[\ref{eomm}], have been simultaneously discretized using the finite time-difference scheme, with $2\times10^5$ iterations and a time step-size of $\Delta t=10^{-3}$. Thermodynamic quantities have been obtained as per Eq.[\ref{workdef},\ref{firstlaw}], and an averaging has been done over $10^4$ cycles of operation (beyond the transient regime). Gaussian random numbers have been obtained from a standard sub-routine, and a zero mean of the corresponding distribution has been ensured. The angular orientation $(\phi)$ has been subjected to periodic boundary conditions between $[0:2\pi]$. The noises in the body frame have been generated from different trails of random numbers to ensure that they are not correlated at any instance. For simulating the corresponding Brownian dynamics, the noise amplitudes are made proportional to $\sqrt{\Delta t}$ to ensure the correlation dictated by the respective FDRs. The temperature $(T)$ to which all the DoF are equilibrated is changed after every half-cycle, as dictated by the instantaneous value of the protocol $k(t)$.

The following values of the system-parameters are considered wherever they are required to be constants, unless otherwise mentioned : $\Gamma_r=1, \Delta\Gamma=9, k_{\phi}=50, \phi_0=0.78, k_\text{max}=2, k_\text{min}=1, T_H=2, T_C=1, \tau=500$ (quasi-static limit). Throughout the simulation, the value of $\Gamma_r$ has been kept equal to the lowest one \cite{mandal2024diffusion} among $\{\Gamma_\parallel,\Gamma_\perp\}$. For practical circumstances, $\Gamma_r$ can vary between the two translational mobilities, depending on the size parameters of the particle \cite{dhont1996introduction}.

Using the above set of parameters, the isotropic values of average extracted work, input heat and quasi-static efficiency (as per Eq.[\ref{isowork},\ref{isoheat},\ref{isoeffi}]) come out to be $-0.693$, $2.386$, and $0.29$ respectively. The corresponding Carnot efficiency is $\eta_C=0.5$.

\subsection{Configuration space}

\begin{figure}[htp]
    \centering
    \includegraphics[width=8cm]{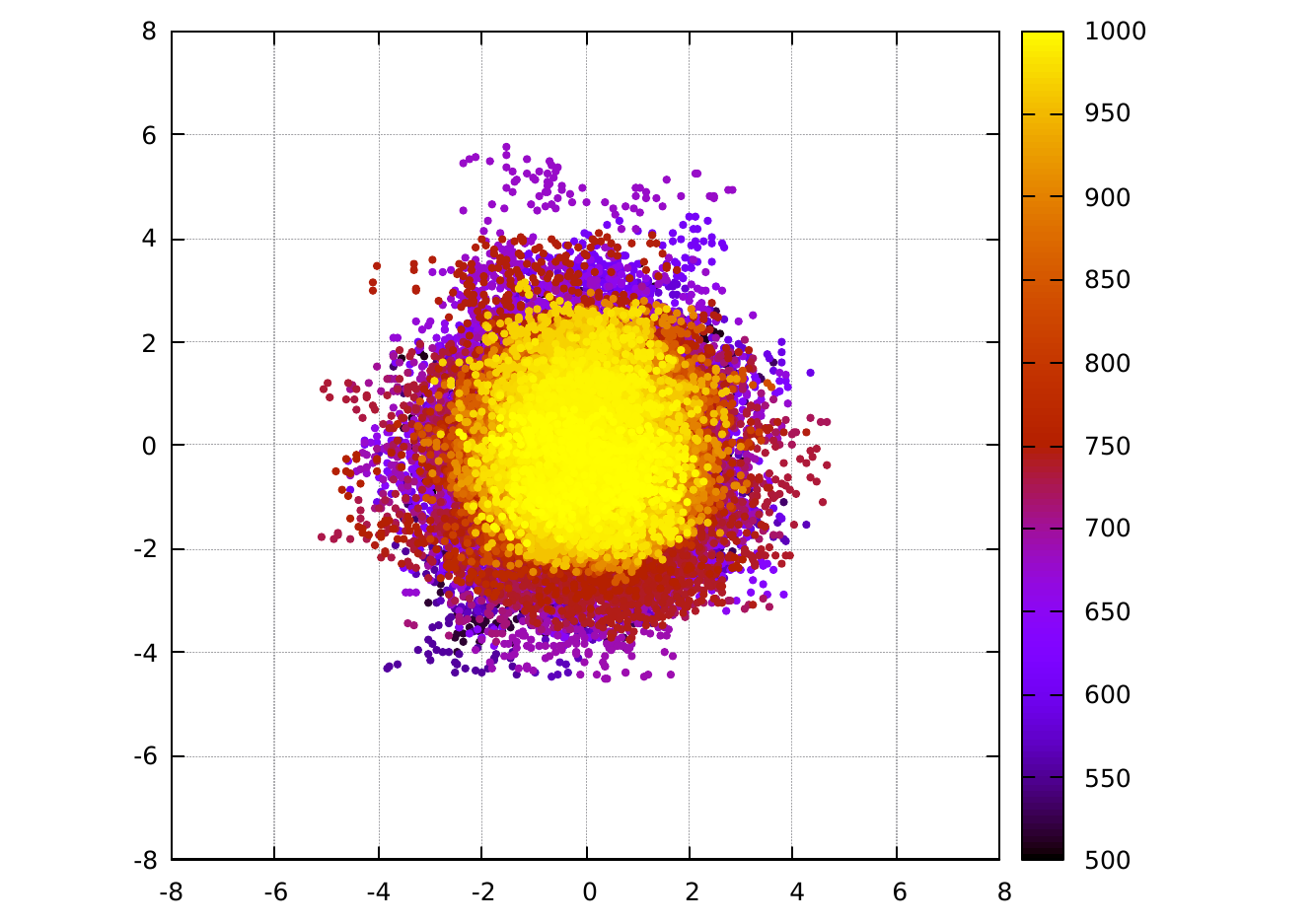}
    \includegraphics[width=8cm]{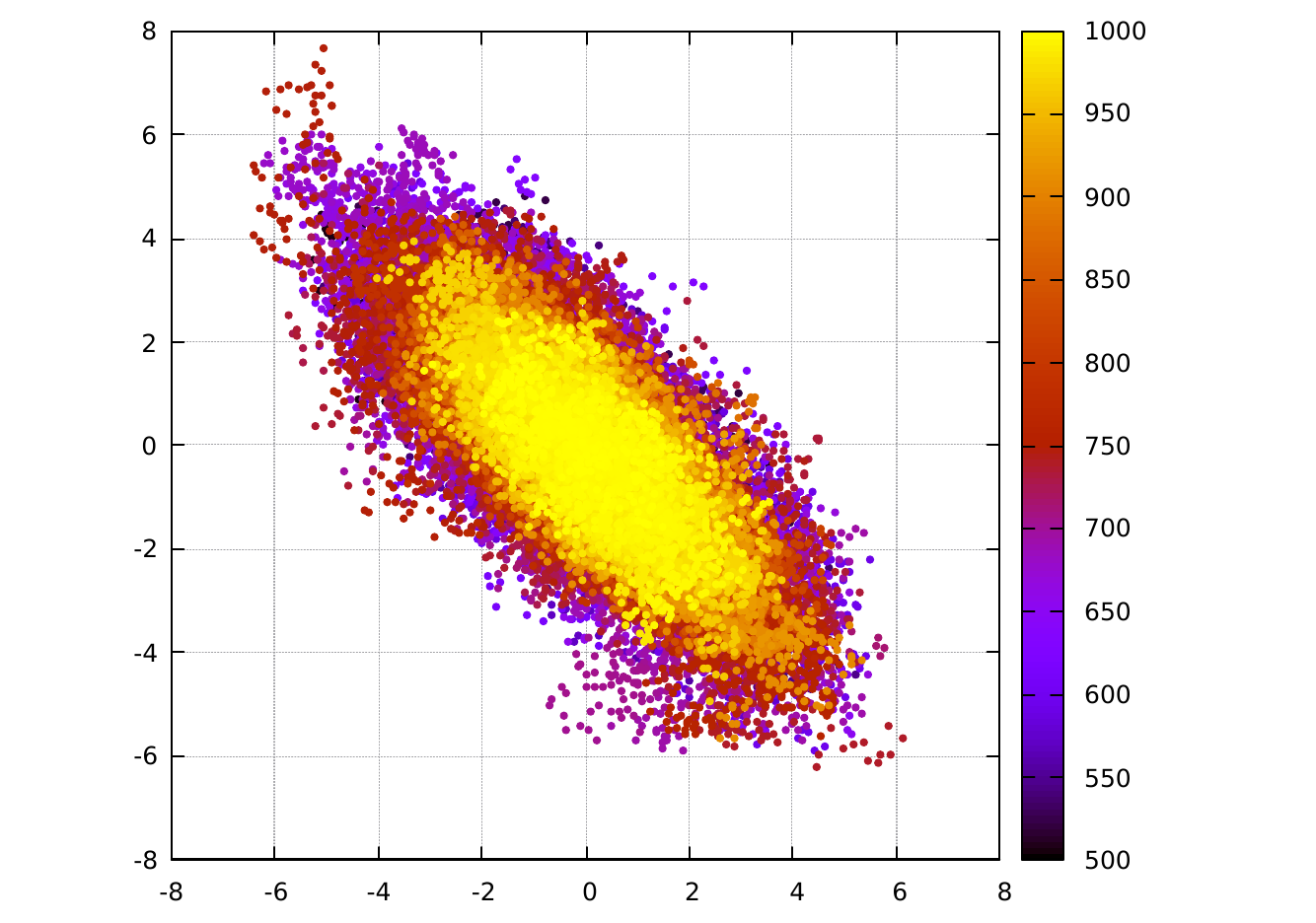}
    \caption{Configuration space of the ellipsoid, subjected to a time-dependent protocol, for $\phi_0=0$ (top) and $\phi_0=0.78$ (bottom) respectively. The color-scale denotes the time during one complete cycle.} 
\label{config}    
\end{figure}

The configuration space (as opposed to a phase space) contains an aggregate of points, where each point signifies the coordinates of center of mass of the ellipsoid, $\{x,y\}$ (at a particular time instance), during its translation in the lab frame (see Fig.[\ref{config}]). As the stiffness constant is linearly increased in the second half of the cycle (that is, the compression phase), the blob of points is seen to be more confined after $t=750$. We see that for $\phi_0=0.78$, the distribution of points in the configuration space is
not symmetric about $\{0,0\}$ (as compared to $\phi_0=0$), and there exists a higher correlation between $x$ and $y$. This can be attributed to the maximal coupling between translational DoF at $\phi_0=0.78$, as dictated by $\Gamma_{xy}$. The distribution of points become symmetric only for minimal/zero coupling. Thus, in the steady state, when the long axis of the ellipsoid is tilted at a mean angle of $\phi_0=0.78$ (w.r.t. the lab frame), it covers a larger span of the configuration space, owing to its highly-correlated translational dynamics. Quantitatively, this feature can be captured via the dynamical cross-correlation between $x$ and $y$ --- another significant moment, besides the variance of position. 

\subsection{Dynamical cross-correlation, $\langle xy \rangle$}

Studying the variation of the cross-moment $(\langle xy \rangle)$ with the torque parameters $(k_\phi,\phi_0)$ essentially reveals the dependence of translational motion of an ellipsoid on its orientational dynamics, besides the dynamical coupling of the DoF. Conversely, this can also be seen as a measure of the correlation between the polar angle $(\theta)$ and the angular orientation $(\phi)$.

\begin{figure}[htp]
    \centering
    \includegraphics[width=8cm]{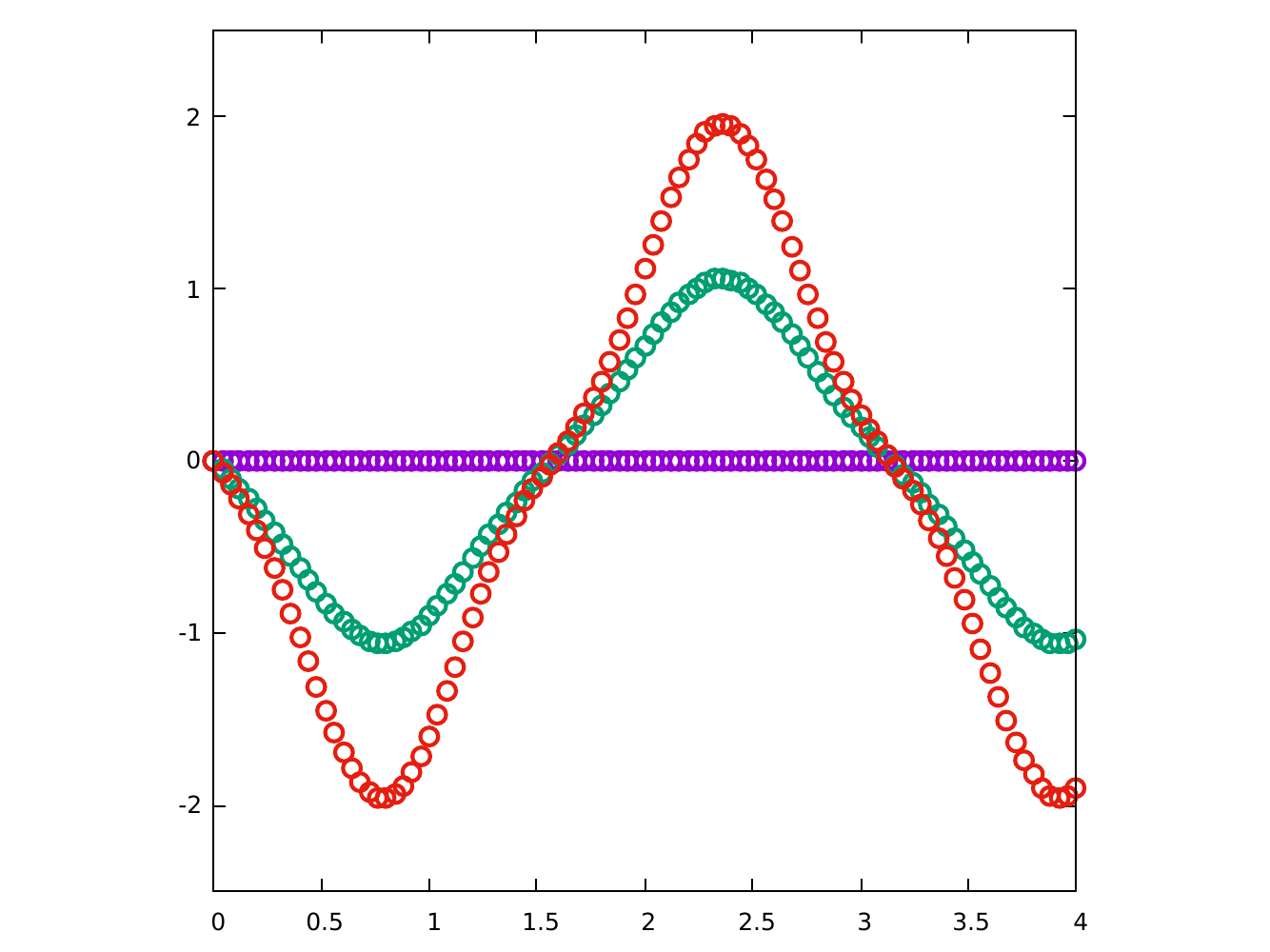}
    \caption{Plot of $\langle xy \rangle$ versus $\phi_0$ (in radian), for various $\Delta\Gamma$ : purple, green and red points are for $\Delta\Gamma=0,4,9$ respectively.} 
\label{corr1}    
\end{figure}

Fig.[\ref{corr1}] shows that there exists no correlation between the DoF, when $\Delta\Gamma=0$ --- for a spherical particle, the dynamical equations are not coupled. However, this correlation shows a periodic variation with $\phi_0$, as soon as the particle is considered to be an ellipsoid. This can be generalized to any anisotropic particle --- the peak correlation at $\phi_0=0.78$ increases when the non-zero value of $\Delta\Gamma$ is increased. One must note that all these variations are steady-state, averaged ones. For $\phi_0=0$, even when the particle is an ellipsoid, the dynamical coupling of the DoF vanishes (on an average), also, the lab and body frames become equivalent. This leads to $\langle xy \rangle=0$. For angles close to $\phi_0=0$, $\langle xy \rangle$ increases linearly, and the periodic variation becomes prominent only in the vicinity of $\phi_0=0.78$. The sign of $\langle xy \rangle$ is flipped after $\phi_0=1.57$ (which is essentially equal to $\frac{\pi}{2}$ radian), that is, when the long axis of the ellipsoid becomes perpendicular to the reference axis in the lab frame. Clearly, the cross-moment depends on $\Gamma_{xy}$. 

\begin{figure}[htp]
    \centering
    \includegraphics[width=8cm]{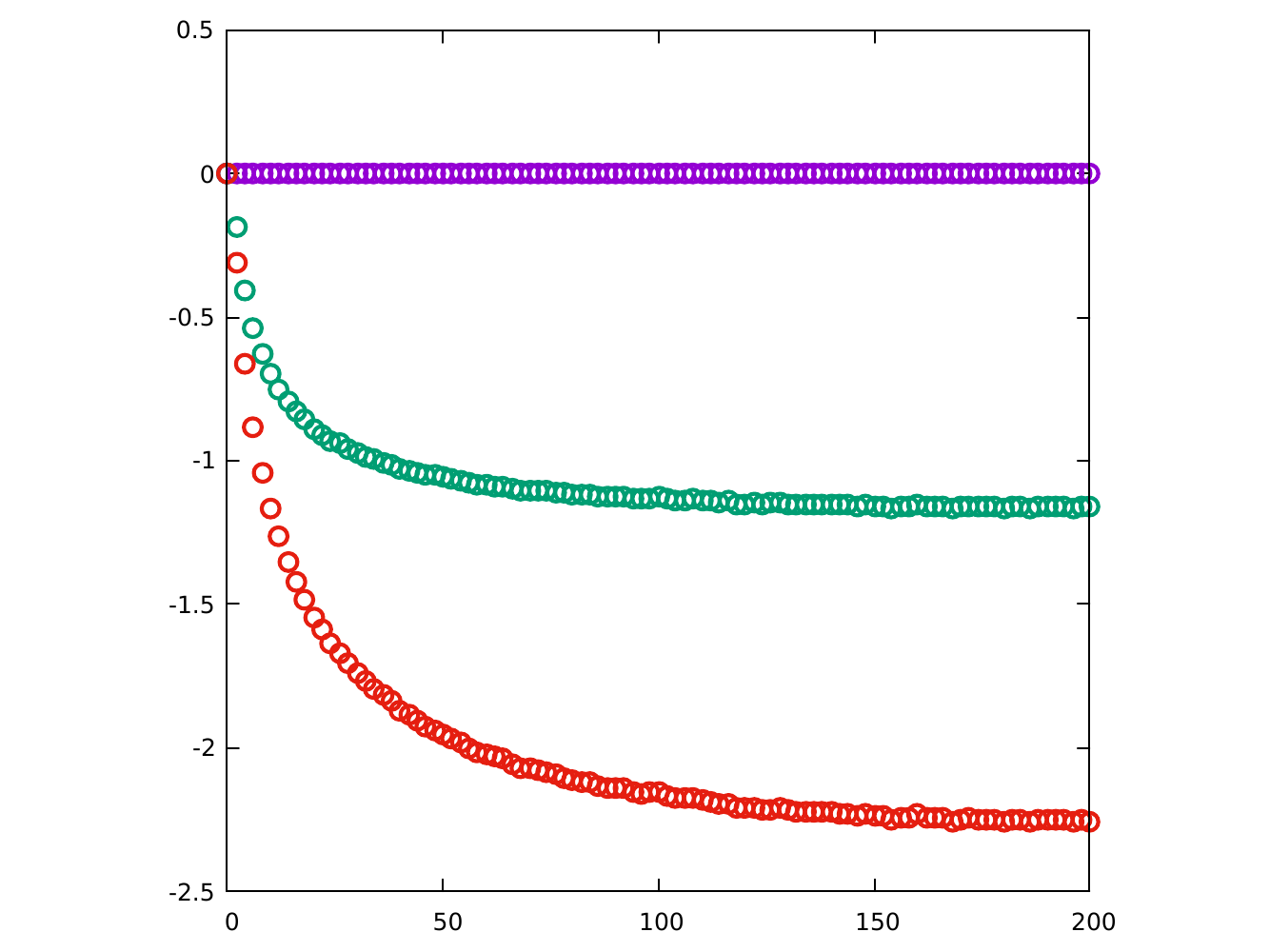}
    \caption{Plot of $\langle xy \rangle$ versus $k_\phi$, for various $\Delta\Gamma$ : purple, green and red points are for $\Delta\Gamma=0,4,9$ respectively.} 
\label{corr2}    
\end{figure}

Fig.[\ref{corr2}] shows the variation of the same with the torque-strength $k_\phi$, at $\phi_0=0.78$ (that is, the regime of maximal coupling). For a sphere, this variation is saturated at zero, and shows no dependence on $k_\phi$ whatsoever. This is indeed expected, as a passive sphere has no specific orientation of its own. For $\Delta\Gamma\neq0$, the value of $\langle xy \rangle$ saturates at its respective peak values at larger values of $k_\phi$, consistent with Fig.[\ref{corr1}]. This happens because when $k_\phi$ is increases, the thermal fluctuations in the dynamics of $\phi$ are suppressed, leading to a decrease in the spread of the corresponding Gaussian distribution (the variance decreases for a given bath temperature) --- that is, more values of $\phi$ resides closer to the mean, which is set at $\phi_0=0.78$ in our case (see Eq.[\ref{phimom}]). Hence, the region of maximal coupling is preferred by the orientational dynamics, when $k_\phi$ is increased. At $k_\phi=0$, however, the angular orientation becomes a freely-diffusing variable, and its mean becomes zero (due to all possible random orientations) and variance diverges at large time. As a result, the correlation between DoF is diminished to zero (on an average). Hence, for the coupling between $x$ and $y$ to survive statistically, the passive ellipsoid must be tilted at a finite, mean orientation w.r.t. to the lab frame --- this is possible only with an externally applied, restoring torque.

\subsection{Average extracted work, $\langle W \rangle$}

\begin{figure}[htp]
    \centering
    \includegraphics[width=8cm]{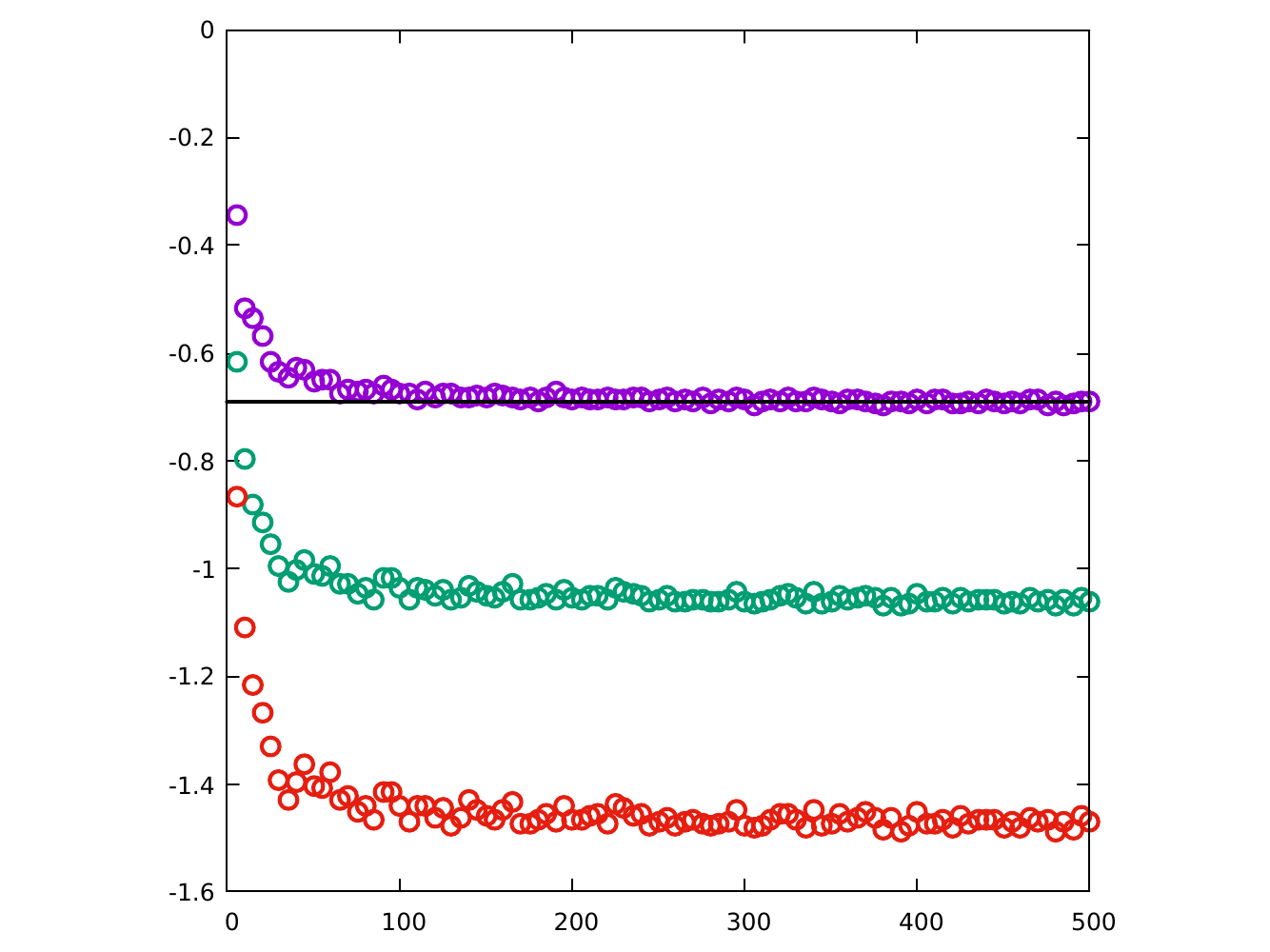}
    \caption{Plot of $\langle W \rangle$ versus $\tau$, for various $\Delta\Gamma$ : purple, green and red points are for $\Delta\Gamma=0,4,9$ respectively. The isotropic value is marked by a solid black line.} 
\label{versustau}    
\end{figure}

In Fig.[\ref{versustau}], we have plotted the average extracted work with the cycle time, for various values of $\Delta\Gamma$. Starting from zero, $\langle W \rangle$ decreases and finally saturates to a negative value for large $\tau$ (quasi-static limit). For the isotropic case, this saturation is consistent with the value obtained analytically. Thus, work can be extracted from the devised stochastic engine, and this remains valid for all values of $\tau$ --- that is, work is not observed to flip its sign. Moreover, the saturation value increases when $\Delta\Gamma$ is increased from zero. This means that the anisotropic case yields more work as compared to the isotropic one. The comparison between the performance of this engine in the two cases will be discussed later. The plots given thereafter are all obtained at the quasi-static limit.

\begin{figure}[htp]
    \centering
    \includegraphics[width=8cm]{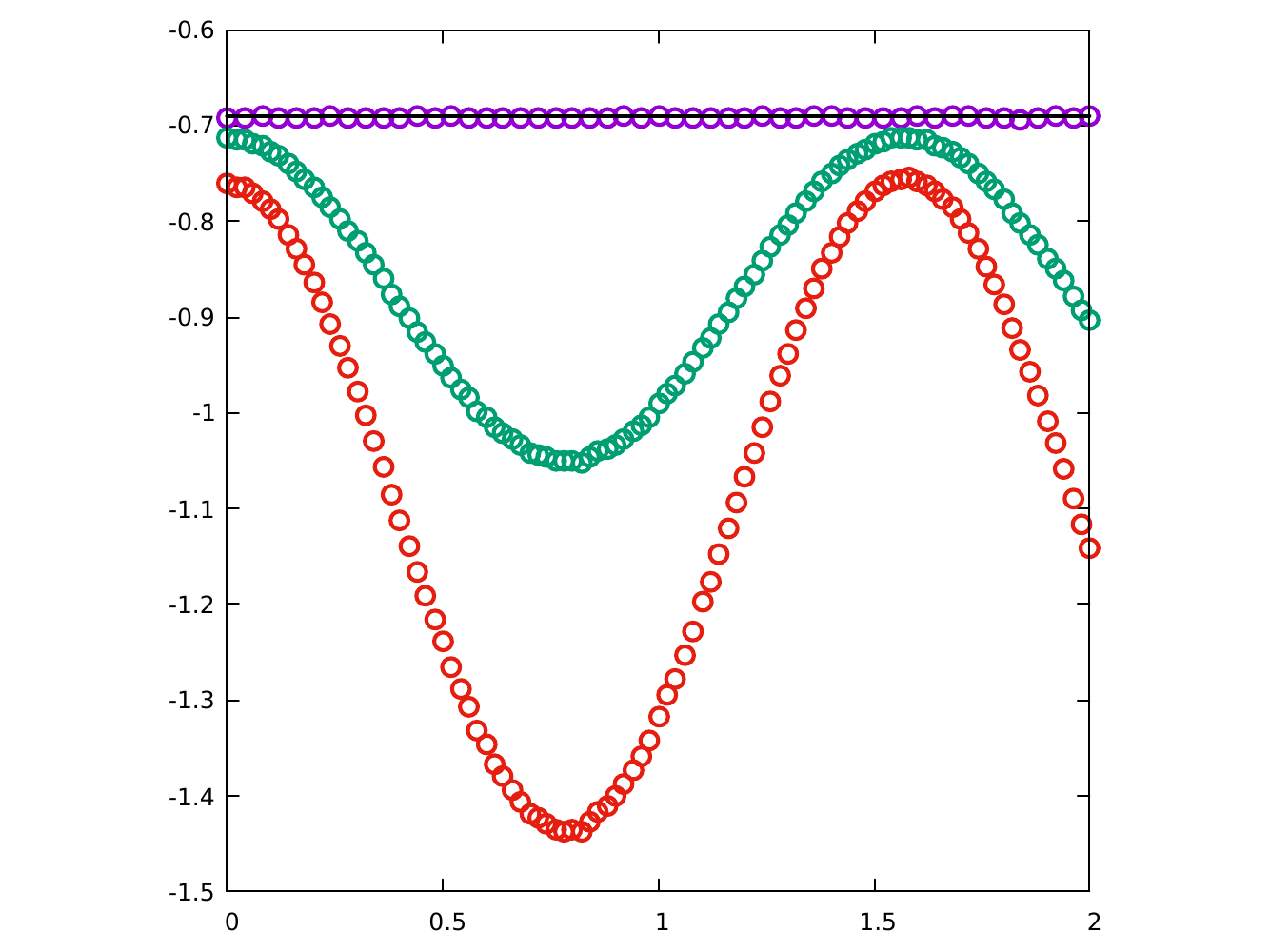}
    \caption{Plot of $\langle W \rangle$ versus $\phi_0$, for various $\Delta\Gamma$ : purple, green and red points are for $\Delta\Gamma=0,4,9$ respectively. The isotropic value is marked by a solid black line.} 
\label{versusphi0}    
\end{figure}

In Fig.[\ref{versusphi0}], we obtain the isotropic work for $\Delta\Gamma=0$, having no variation with $\phi_0$, as expected. The cases with $\Delta\Gamma\neq0$ become closer to the isotropic limit at $\phi_0=0$. This merging can become more prominent at large $k_\phi$, when the values of $\phi$ will reside closer to $\phi_0=0$ (this will become evident in the subsequent plots). One must note here that the working substance still has an anisotropic shape, even when $\phi_0=0$, and Eq.[\ref{isowork}] remains valid only when the dynamical coupling is completely lost (which clearly does not happen sufficiently with $k_\phi=50$). For $\phi_0=0.78$, however, the anisotropic cases become distinctly different from the isotropic value, such that the peak becomes proportional to $\Delta\Gamma(\neq 0)$. These variations are repeated periodically. Hence, $\Gamma_{xy}$ clearly governs the extracted work, besides other parameters.

\begin{figure}[htp]
    \centering
    \includegraphics[width=8cm]{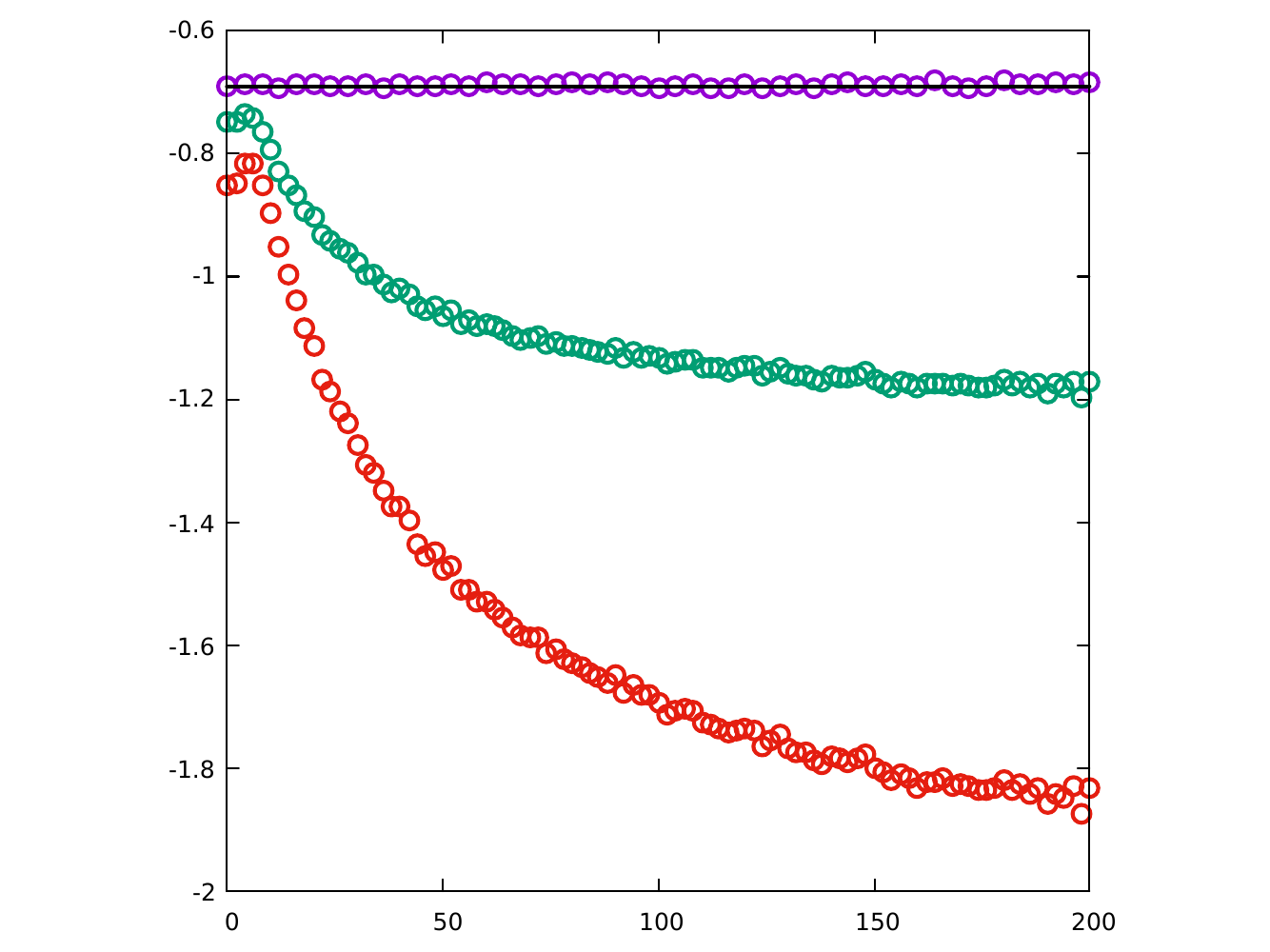}
    \caption{Plot of $\langle W \rangle$ versus $k_\phi$, for various $\Delta\Gamma$ : purple, green and red points are for $\Delta\Gamma=0,4,9$ respectively. The isotropic value is marked by a solid black line.} 
\label{versuskphi}    
\end{figure}

As shown in Fig.[\ref{versuskphi}], the isotropic limit has no variation with $k_\phi$, as expected. For $\phi_0=0.78$, the isotropic cases saturate to their respective peak values at larger $k_\phi$, due to reasons mentioned above. When $k_\phi=0$ and $\Delta\Gamma\neq0$, the orientation of the ellipsoid shows free diffusion, and as a result, the system is not governed by Eq.[\ref{eom3}]. Hence, the magnitude of work obtained in this case differs from the isotropic limit. As soon as $k_\phi$ increases slightly from zero, the restoring torque immediately tries to bring the orientation to a non-zero mean value (which is set in the MCR). The small peak occurs due to this sudden change in the value of $k_\phi$, leading to a completely different dynamics of $\phi$ when the torque is switched on. Hence, as the torque is switched on, the value of the extracted work increases a little (closely up to the isotropic bound) but then rapidly decreases to saturate at the respective anisotropic values, when $k_\phi$ starts to take larger values and maximal coupling sets in the system due to the chosen value of $\phi_0$. This leads to be distinct separation of the anisotropic cases from the isotropic bound.

\begin{figure}[htp]
    \centering
    \includegraphics[width=8cm]{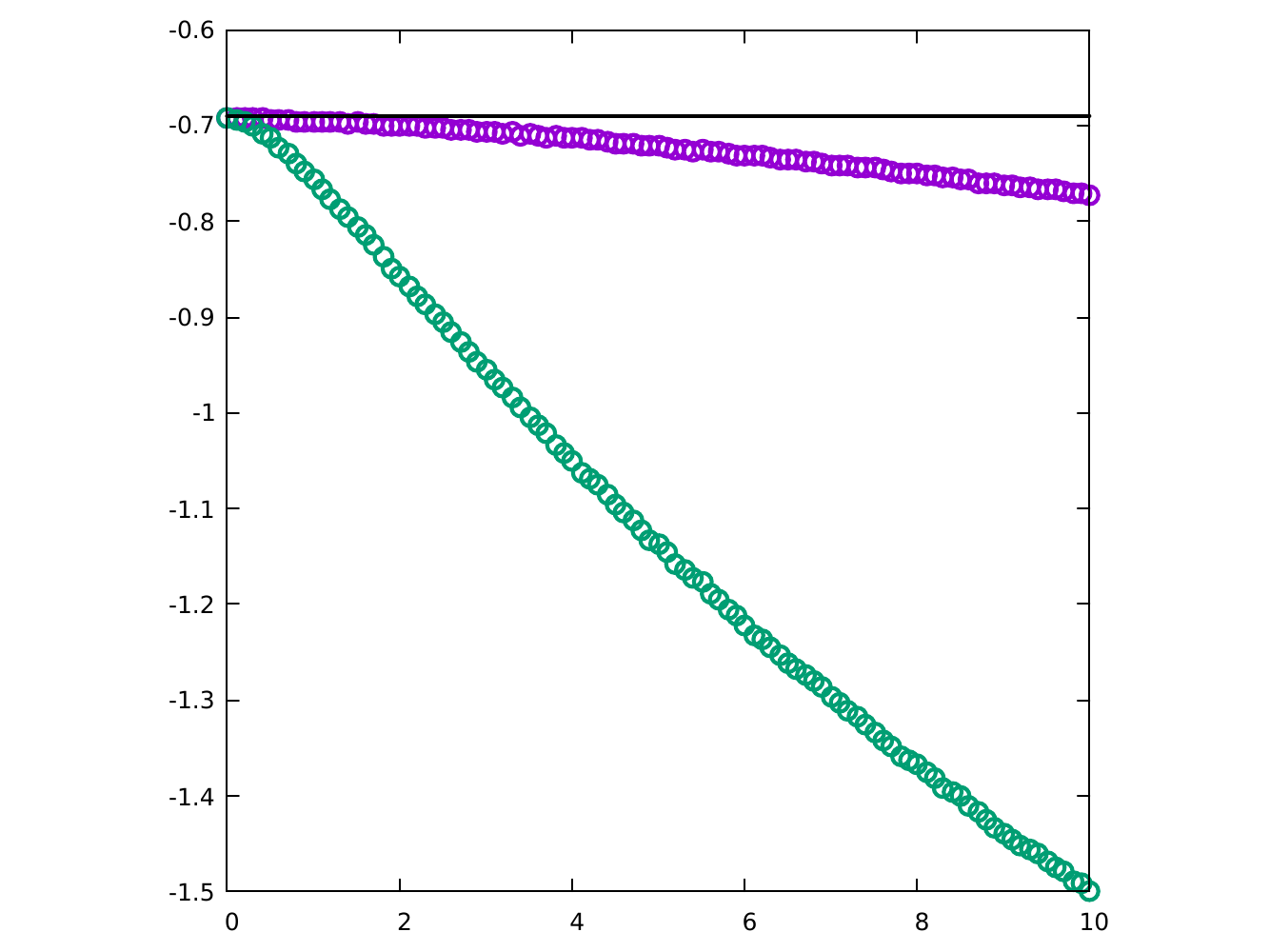}
    \caption{Plot of $\langle W \rangle$ versus $\Delta\Gamma$, for various $\phi_0$ : purple and green points are for $\phi_0=0,0.78$ respectively. The isotropic value is marked by a solid black line.} 
\label{versusdelta1}    
\end{figure}

\begin{figure}[htp]
    \centering
    \includegraphics[width=8cm]{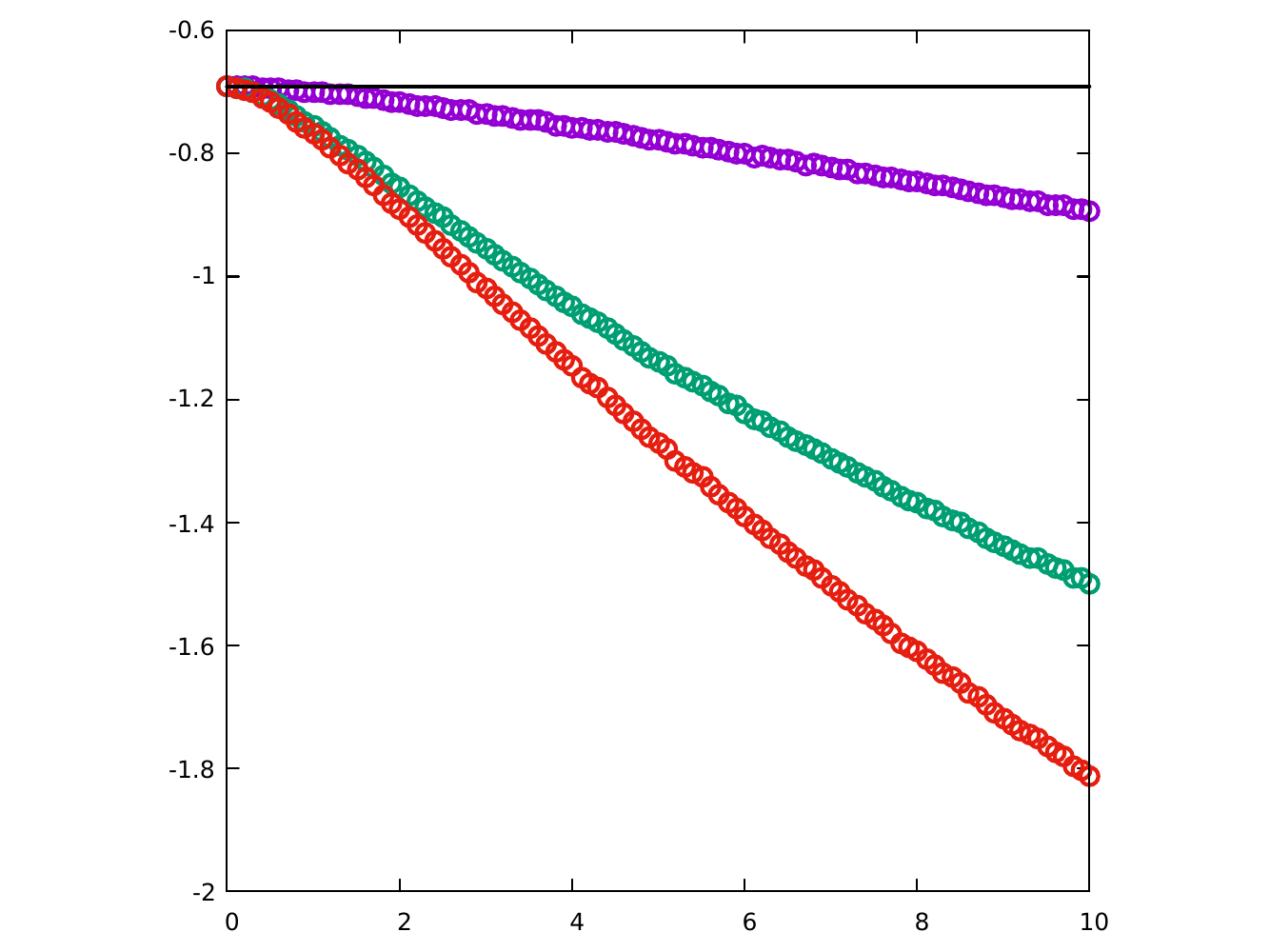}
    \caption{Plot of $\langle W \rangle$ versus $\Delta\Gamma$, for various $k_\phi$ : purple, green and red points are for $k_\phi=1,50,200$ respectively. The isotropic value is marked by a solid black line.} 
\label{versusdelta2}    
\end{figure}

In Fig.[\ref{versusdelta1}], we see that the average extracted work increases to larger negative values with an increase in $\Delta\Gamma$. Starting from the isotropic bound, this increase is linear away from this bound --- the small non-linear region around $\Delta\Gamma=0$ will be studied in the later sections. However, the slope of this linear variation is much less for $\phi_0=0$ as compared to the MCR. This causes it to be closer to the isotropic limit. By increasing $k_\phi$, this proximity can be increased. Nonetheless, the system yields widely-separated work outputs corresponding to the minimal and maximal coupling regimes. Keeping $\phi_0$ fixed at $0.78$, the effect of $k_\phi$ on the slope of the aforesaid linear variation is studied in Fig.[\ref{versusdelta2}]. For reasons mentioned earlier, a higher $k_\phi$ will enable the system to yield a higher anisotropic work, as compared to a lower value of $k_\phi$. Conversely, a larger $k_\phi$ ensures a greater slope of the linear variation in the $\langle W \rangle - \Delta\Gamma$ plot. Thus, for larger values of the mobility difference, the average extracted work varies in proportion to it, that is, work output can be enhanced from the isotropic case if the working substance shows more anisotropy in its shape. This is true for any bi-axial particle showing a dichotomy in its friction profile.

\begin{figure}[htp]
    \centering
    \includegraphics[width=8cm]{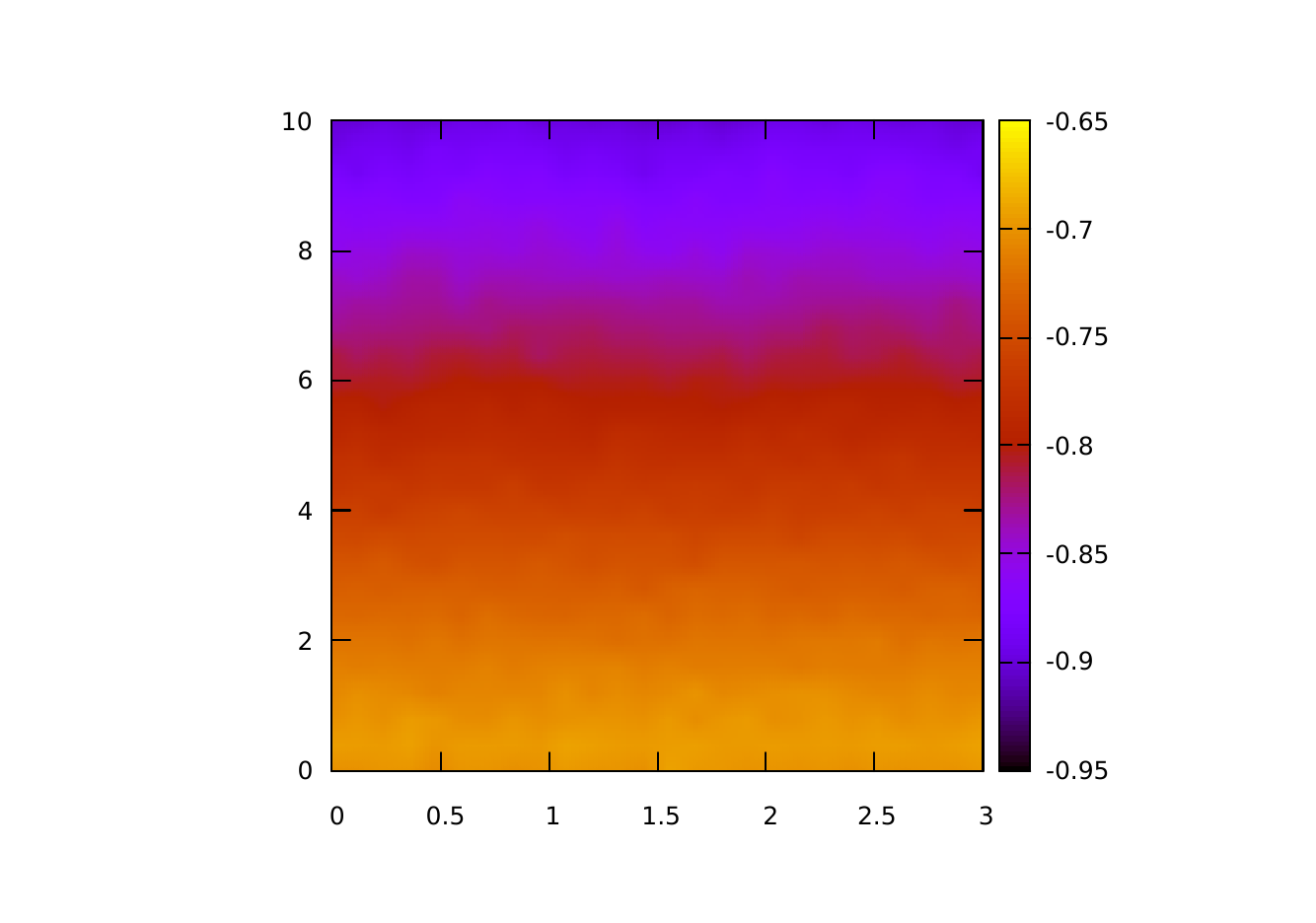}
    \includegraphics[width=8cm]{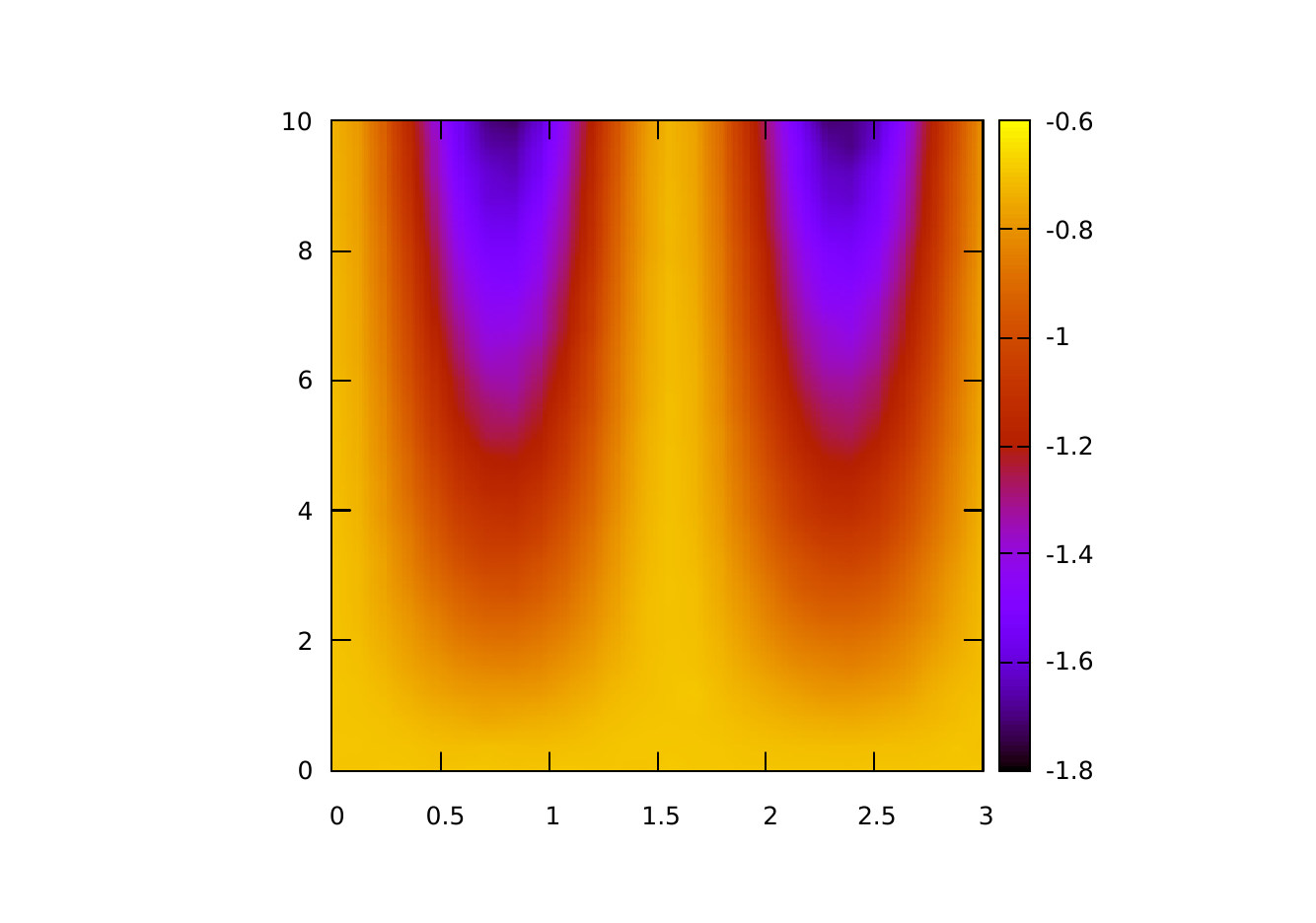}
    \caption{Average extracted work, for various $k_\phi$ : $\phi_0$ along X-axis, $\Delta\Gamma$ along Y-axis, $\langle W \rangle$ along color-scale --- (top) $k_\phi=1$, (bottom) $k_\phi=100$.} 
\label{phase1}    
\end{figure}

\begin{figure}[htp]
    \centering
    \includegraphics[width=8cm]{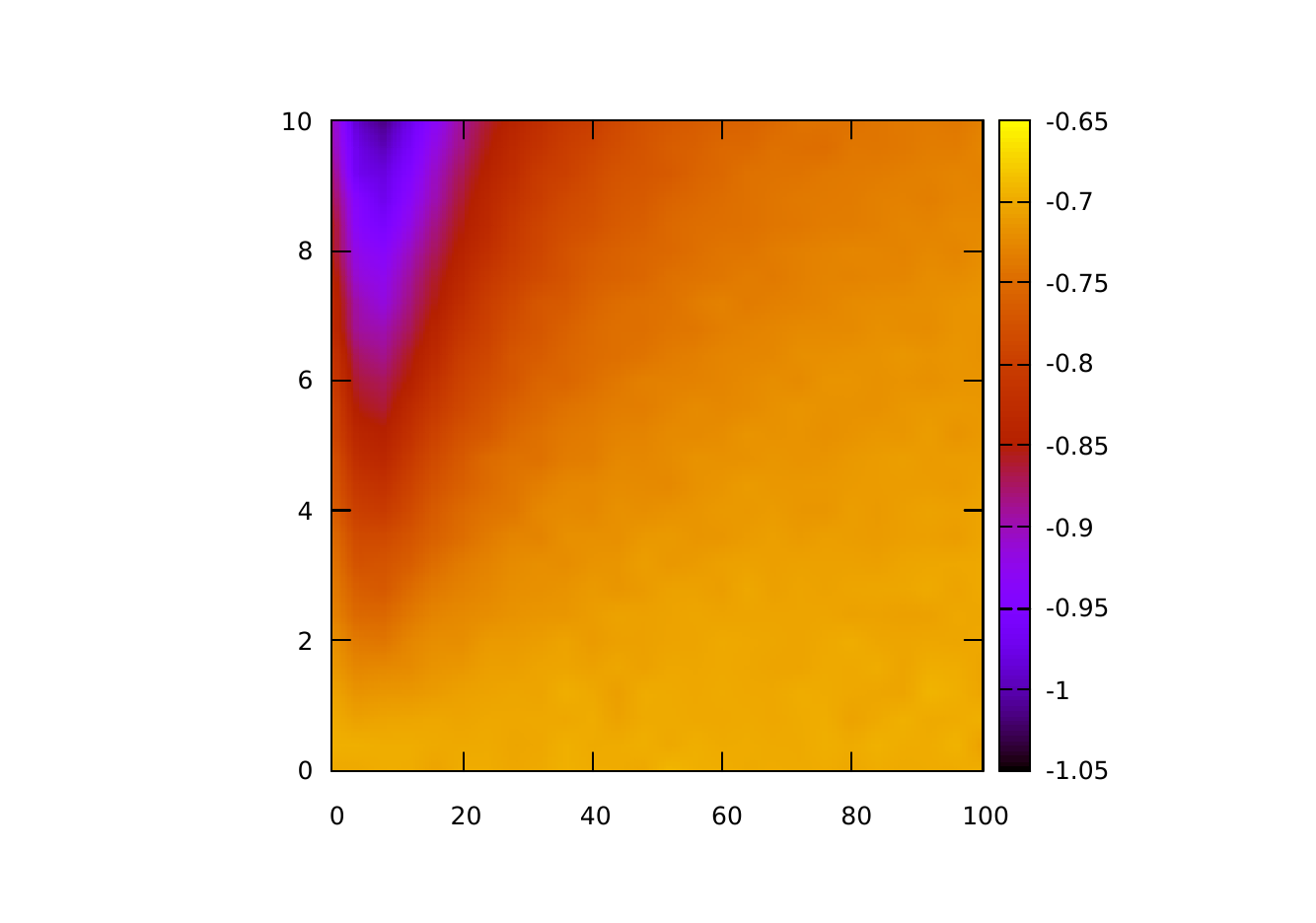}
    \includegraphics[width=8cm]{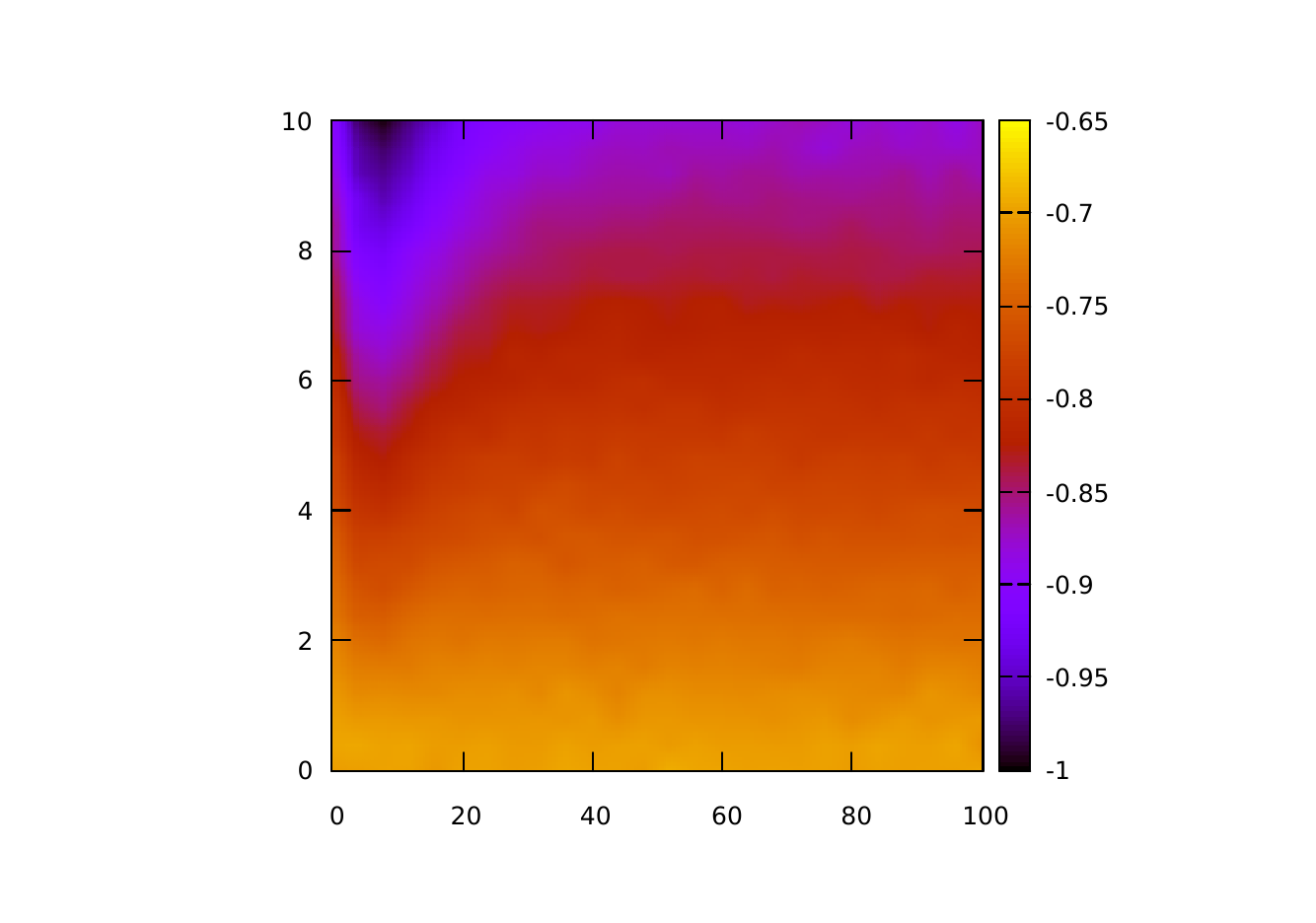}
    \includegraphics[width=8cm]{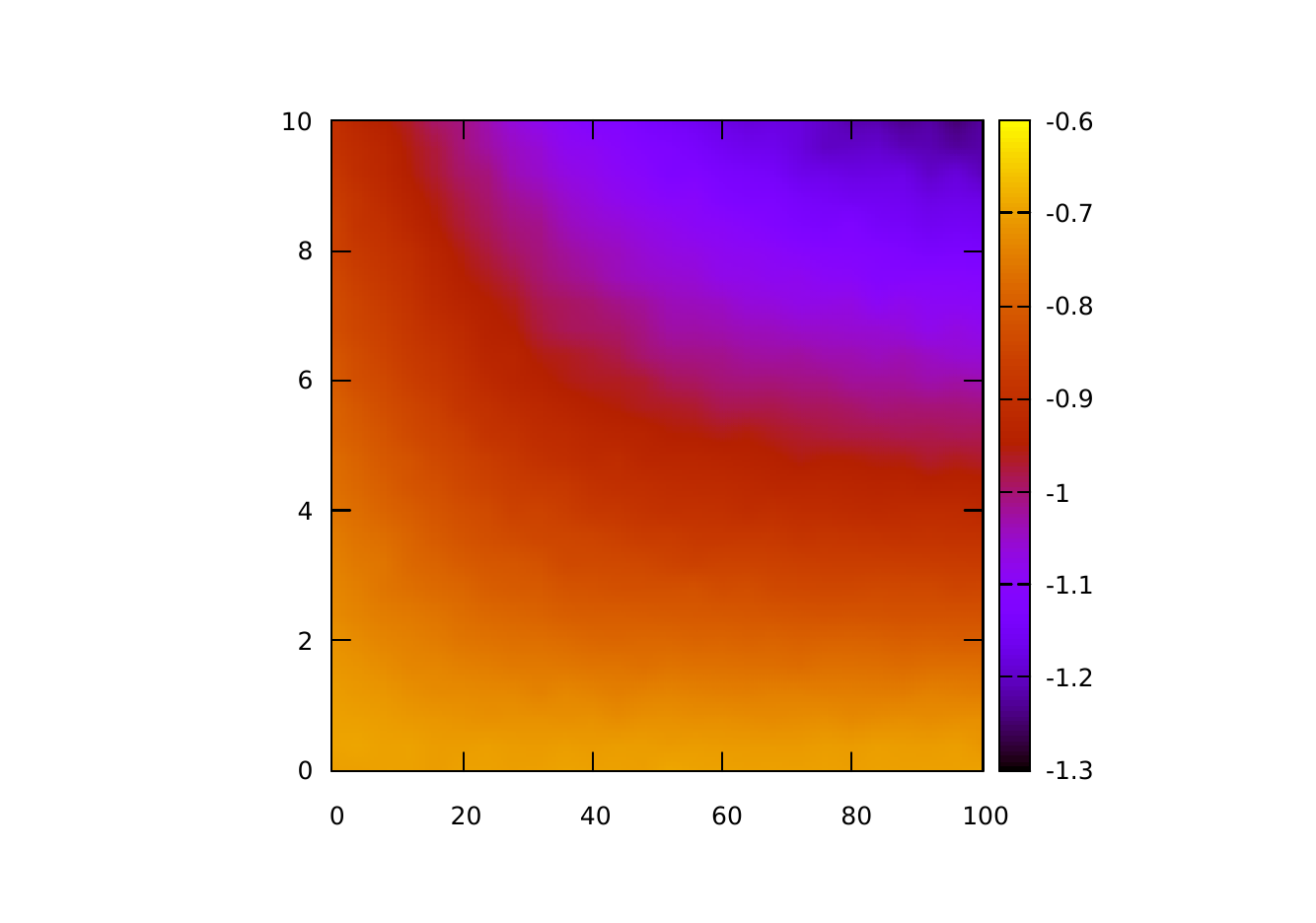}
    \includegraphics[width=8cm]{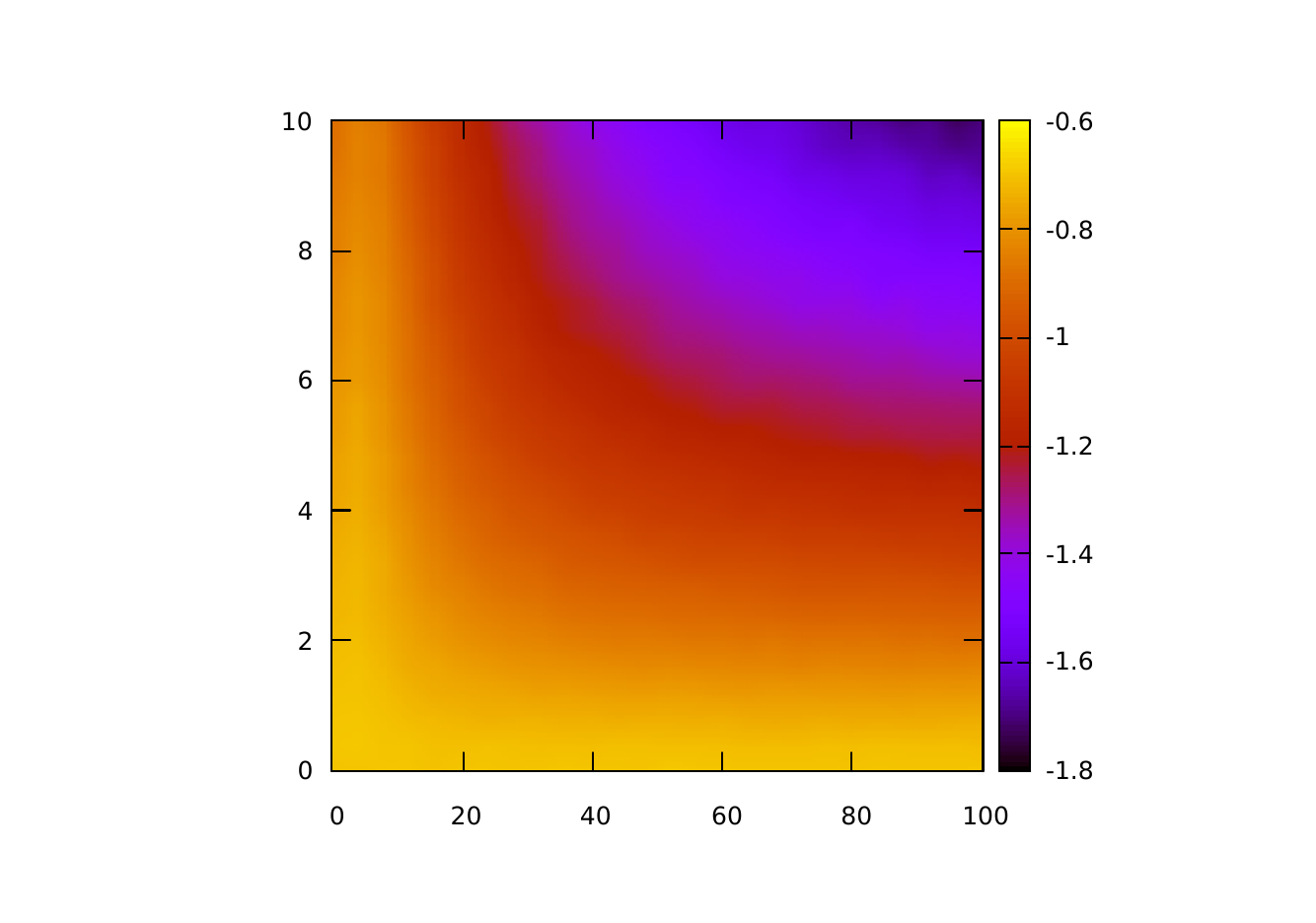}
    \caption{Average extracted work, for various $\phi_0$ : $k_\phi$ along X-axis, $\Delta\Gamma$ along Y-axis, $\langle W \rangle$ along color-scale --- (top to bottom) $\phi_0=0,0.2,0.4,0.78$.} 
\label{phase2}    
\end{figure}

Having discussed all the causative parameters of extracted work, it becomes significant to visualize all the variations simultaneously in a phase diagram. Here, the objectives are two-fold : first, the phase diagrams will enable us to clearly understand the inter-relationship between the shape parameter $(\Delta\Gamma)$ and the external torque parameters $(k_\phi,\phi_0)$ and how they affect work at the expense of one another; secondly, the diagrams will assist in identifying regions where the system acts as an isotropic Stirling engine and where the anisotropy has intruded --- we will try to clearly demarcate these two regions and perhaps try to understand how the system undergoes a transition between these regions. In Fig.[\ref{phase1}], extracted work shows no variation with $\phi_0$ for a lower value of the torque. In this region, only the mobility difference acts as the deciding factor, where $\Delta\Gamma=0$ sets the isotropic limit in the system and the system increasingly deviates from this limit as $\Delta\Gamma$ is increased. As soon as the torque is taken to higher values, the orientation gets restored to a finite value. Now, $\langle W \rangle$ shows a periodic variation with $\phi_0$, the isotropic and anisotropic regions being clearly situated at $0$ and $0.78$ respectively. Not to forget the fact that $\Delta\Gamma\neq0$ also decides whether the system will enter into the anisotropic zone or not --- moving parallel to the Y-axis, at $\phi_0=0.78$, the anisotropic work increases in magnitude with an increase in $\Delta\Gamma$. In Fig.[\ref{phase2}], the phase diagrams have been obtained for various $\phi_0$. For $\phi_0=0$, the entire parameter-space is predominantly affected by the isotropic regime, except for small $k_\phi$ and large $\Delta\Gamma$ --- as soon as the torque is switched on, the extracted work attains a considerable magnitude close to the isotropic limit, and that too, at higher values of $\Delta\Gamma$. As $\phi_0$ is increased a little, the anisotropic regime starts to dominate and we see a saturation of the corresponding work, at higher values of $k_\phi$. The saturation occurs after a small dip, due to a sudden change in the orientational dynamics that occur after an abrupt introduction of the restoring torque (as discussed earlier). Now, if $\phi_0$ is increased further up to the maximal coupling region, the anisotropic work starts to occupy the entire parameter-space, with saturation occurring at higher $k_\phi$. Also, this saturation value increases with $\Delta\Gamma$. At $\phi_0=0.78$, we see that the isotropic regime resides at the region demarcated by $k_\phi=0, \Delta\Gamma=0$---leading to distinctly-separated phases of working of the Stirling engine.

\subsection{Input heat, $\langle Q_1 \rangle$}

\begin{figure}[htp]
    \centering
    \includegraphics[width=8cm]{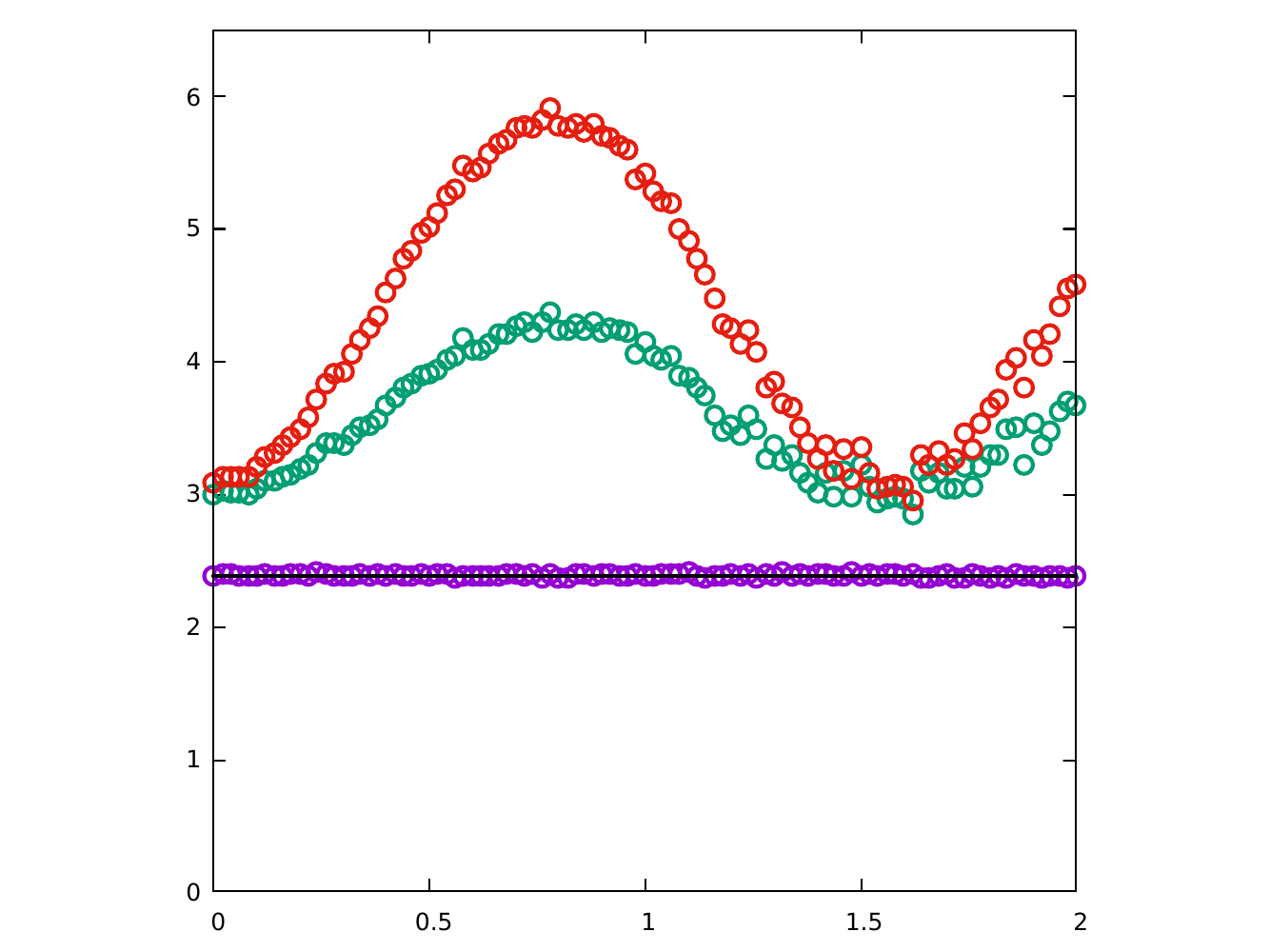}
    \caption{Plot of $\langle Q_1 \rangle$ versus $\phi_0$, for various $\Delta\Gamma$ : purple, green and red points are for $\Delta\Gamma=0,4,9$ respectively. For the isotropic case $(\Delta\Gamma=0)$, $k_\phi=0$; and for $\Delta\Gamma\neq0$, $k_\phi=50$. The isotropic value is marked by a solid black line.} 
\label{heatversusphi0}    
\end{figure}

Fig.[\ref{heatversusphi0}] shows the variation of average heat that goes into the system during the expansion phase, with the mean orientation of the working substance --- which is a single Brownian ellipsoid. The isotropic value of this heat is obtained for $\Delta\Gamma=0$, and its shows no variation with the mean orientation, as expected. The anisotropic value peaks at the maximal coupling region $(\phi_0=0.78)$, and the peak values are in proportion with the values of $\Delta\Gamma$. For various $\Delta\Gamma\neq0$ but $\phi_0=0$, the values of heat coincide because of two reasons --- first, the dynamical coupling has vanished (on an average) and the translational dynamics will acquire isotropic features; second, the cost in internal energy (due to the restoring torque) still survives due to a non-zero value of $k_\phi$, which we have kept same for all the cases of $\Delta\Gamma\neq0$. One can also notice that, at $\phi_0=0$, the values of heat for the isotropic benchmark and the (coinciding) anisotropic cases differ by a finite magnitude --- this difference is attributed to the finite change in the average internal energy (due to the external torque) that survives only in case of an ellipsoid. This feature was not manifested in the variation of extracted work, due to obvious reasons. However, the periodic variation with $\phi_0$ is similar for both the thermodynamic quantities.

\begin{figure}[htp]
    \centering
    \includegraphics[width=9.5cm]{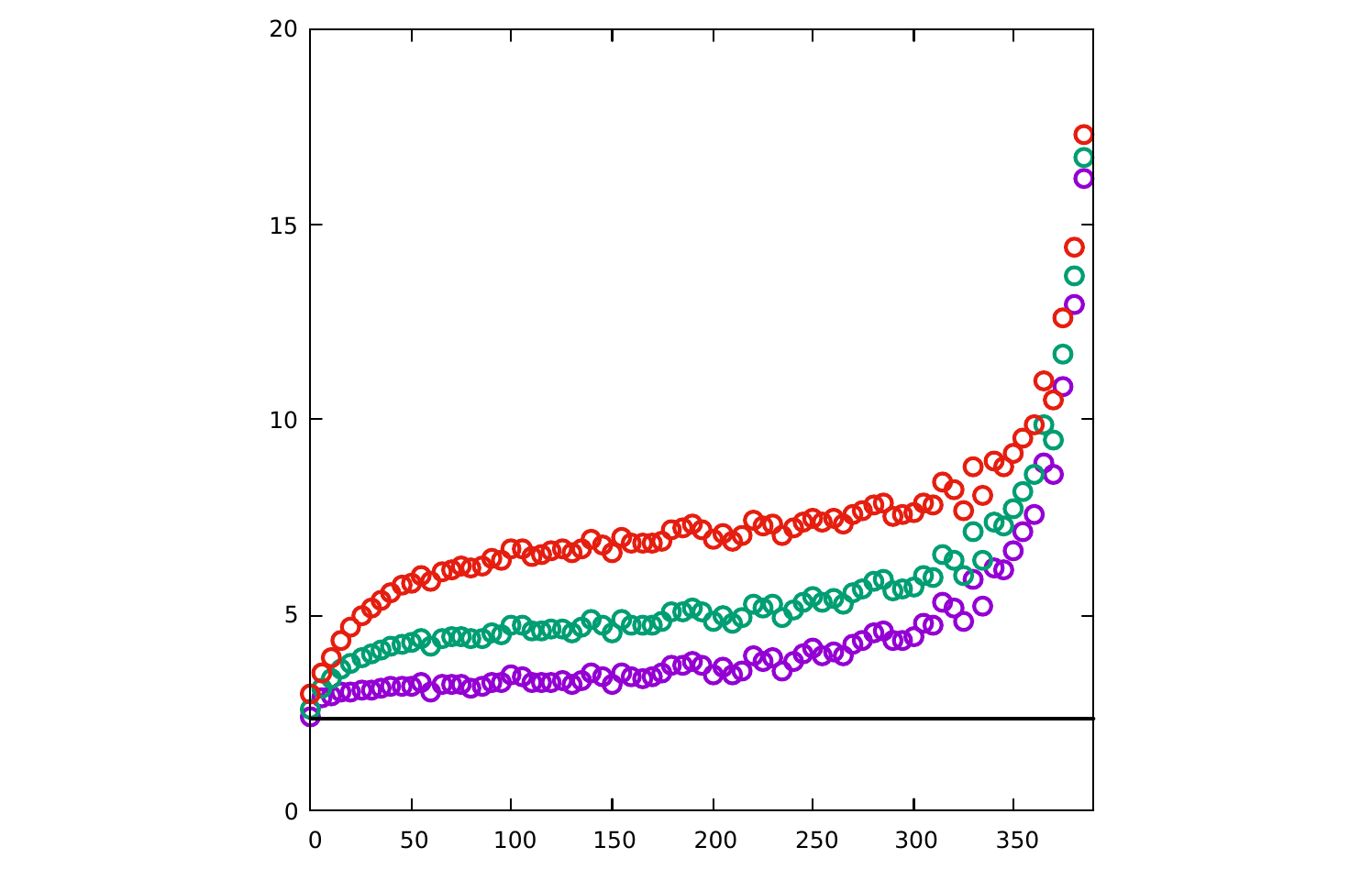}
    \caption{Plot of $\langle Q_1 \rangle$ versus $k_\phi$, for various non-zero $\Delta\Gamma$ : purple, green and red points are for $\Delta\Gamma=1,4,9$ respectively. The isotropic value is marked by a solid black line.} 
\label{heatversuskphi}    
\end{figure}

For $k_\phi=0$ but $\Delta\Gamma\neq0$, the ellipsoid has a freely-diffusing angular orientation, with a zero mean (at all times) and a diverging variance (at large time). As a result, the dynamical coupling vanishes (on an average), besides the internal energy due to the restoring torque being zero. This apparently leads to the isotropy of the system (imitating the behavior of a spherical particle), even when the working substance is an ellipsoid --- this is confirmed by Fig.[\ref{heatversuskphi}]. In such a situation, the dynamics of $\phi(t)$ will only be affected by thermal fluctuations and the ellipsoid acquires all possible random orientations with no specific bias (like a sphere), which inhibits the onset of anisotropy in the system. As $k_\phi$ is increased (in the midst of the MCR), the respective anisotropic heats tend to achieve their distinct peak values. In the above plot, this can be seen as an almost flat region, after a linear increase at smaller $k_\phi$. When $k_\phi$ is increased further to large values, the internal energy due to the restoring torque (which is set at the MCR, for all the anisotropic cases) dominates over the expense of internal energy due to the isotropic harmonic confinement (that is, $k_\phi>>k_\text{max},k_\text{min}$). This leads to the convergence of all the anisotropic cases, irrespective of $\Delta\Gamma$. In the plot, the region after $k_\phi=350$ is the torque-dominated regime. This feature was not observed in case of extracted work, as the torque had no explicit effect as a thermodynamic quantity. Hence, for an extremely high torque, the system absorbs a large amount of heat, but that is not proportionally manifested in the work output. This can be detrimental to the performance of the engine, as will be discussed in the next sub-section.

\begin{figure}[htp]
    \centering
    \includegraphics[width=8cm]{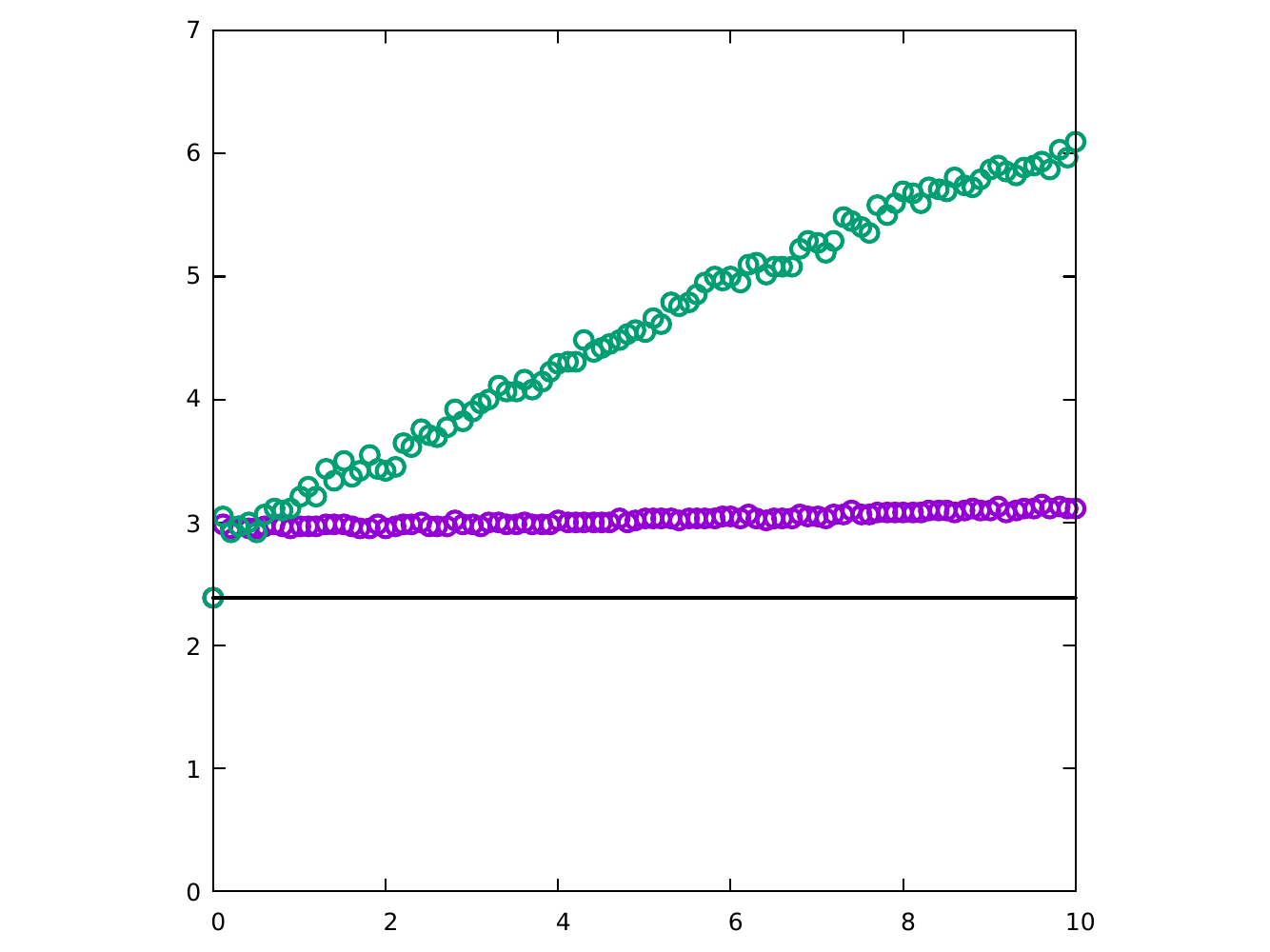}
    \caption{Plot of $\langle Q_1 \rangle$ versus $\Delta\Gamma$, for various $\phi_0$ : purple and green points are for $\phi_0=0,0.78$ respectively. For the isotropic case $(\Delta\Gamma=0)$, $k_\phi=0$; and for $\Delta\Gamma\neq0$, $k_\phi=50$. The isotropic value is marked by a solid black line.} 
\label{heatversusdelta1}    
\end{figure}

\begin{figure}[htp]
    \centering
    \includegraphics[width=8cm]{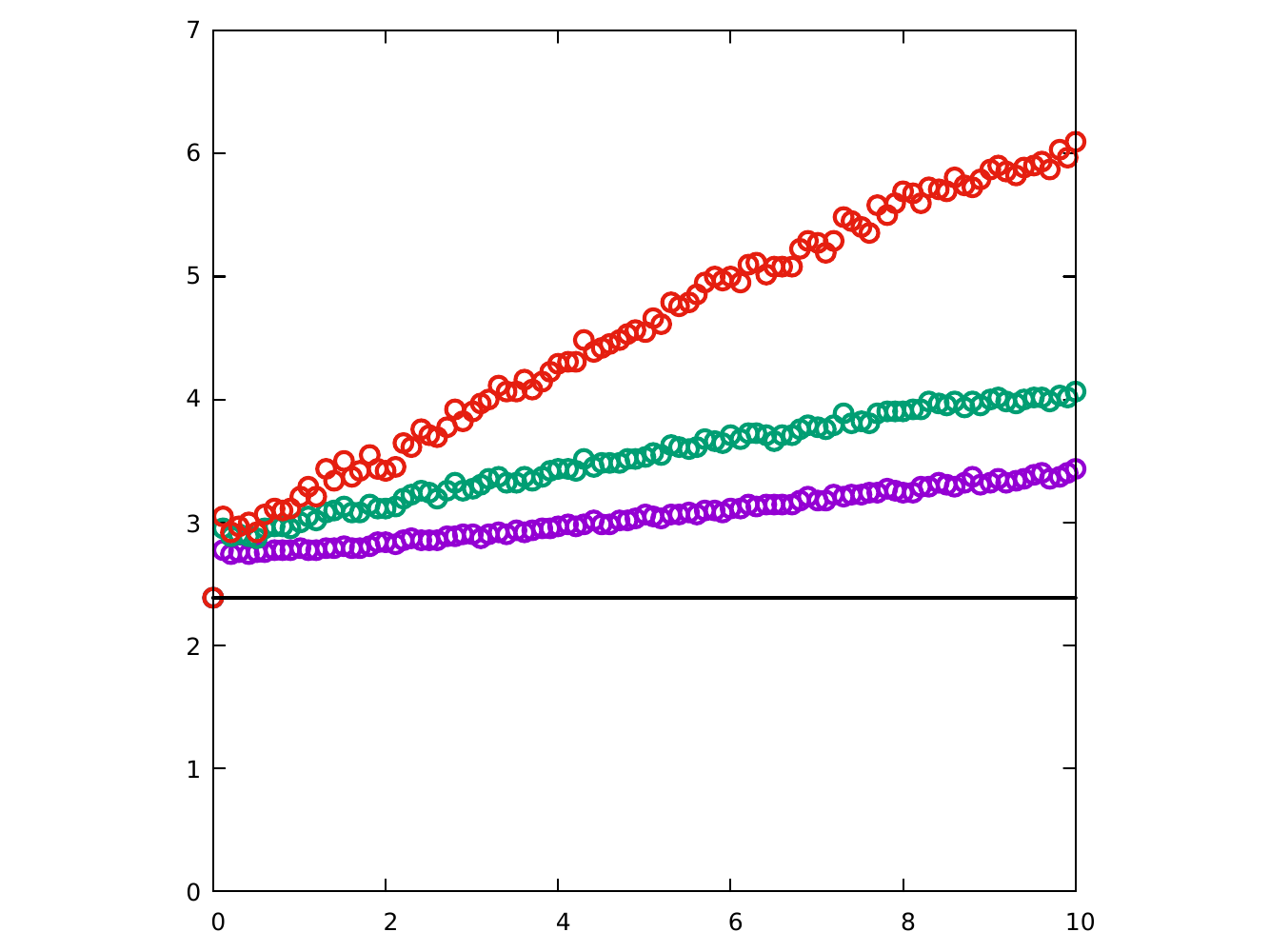}
    \caption{Plot of $\langle Q_1 \rangle$ versus $\Delta\Gamma$, for various non-zero $k_\phi$ : purple, green and red points are for $k_\phi=1,10,50$ respectively. For the isotropic case $(\Delta\Gamma=0)$, however, $k_\phi$ is set to zero. The isotropic value is marked by a solid black line.} 
\label{heatversusdelta2}    
\end{figure}

For $\Delta\Gamma=0$, the cases for various $\phi_0$ coincide with the isotropic value, as shown in Fig.[\ref{heatversusdelta1}]. As soon as the working substance is considered to be an ellipsoid instead of a sphere, the restoring torque comes into play, and the value of heat becomes distinctly different from the isotropic benchmark --- the finite difference between the two is due to the internal energy coming from the torque. Now, for $\phi_0=0$, the input heat shows no variation with $\Delta\Gamma\neq0$, whereas in the MCR, there occurs a linear increase with $\Delta\Gamma\neq0$ --- that is, more the difference in the translational mobilities, more will be the heat absorbed by the system from the hot bath. The same variation is studied in Fig.[\ref{heatversusdelta2}], but for various non-zero values of the torque-strength (in the MCR). In doing so, we have only considered the orientational dynamics of an ellipsoid, subjected to the restoring torque (as opposed to free diffusion), keeping in mind that the isotropic benchmark is satisfied as well. As $\Delta\Gamma$ attains non-zero values, the slope of the linear variation increases with $k_\phi$, with the case for lowest $k_\phi$ being closest to the isotropic limit (for reasons discussed earlier). This is expected since a larger torque will cause a larger expense in the corresponding internal energy, leading to a greater magnitude of absorbed heat. Also, the initiation of the $k_\phi\neq0$ cases shows a finite difference from the isotropic case, owing to this internal energy.

Having studied the variations of extracted work and absorbed heat in the anisotropic regime, with the system parameters involved (mainly shape and orientation), it is an important juncture to judge the performance of this engine in terms of reproducibility of the average work in the midst of several cycles of operation along with its efficiency in yielding so. The following sub-section discusses these aspects and puts forward certain ways of improving the engine performance.

\subsection{Performance of the anisotropic Stirling engine}

\subsubsection{Fluctuation in extracted work}

\begin{figure}[htp]
    \centering
    \includegraphics[width=8cm]{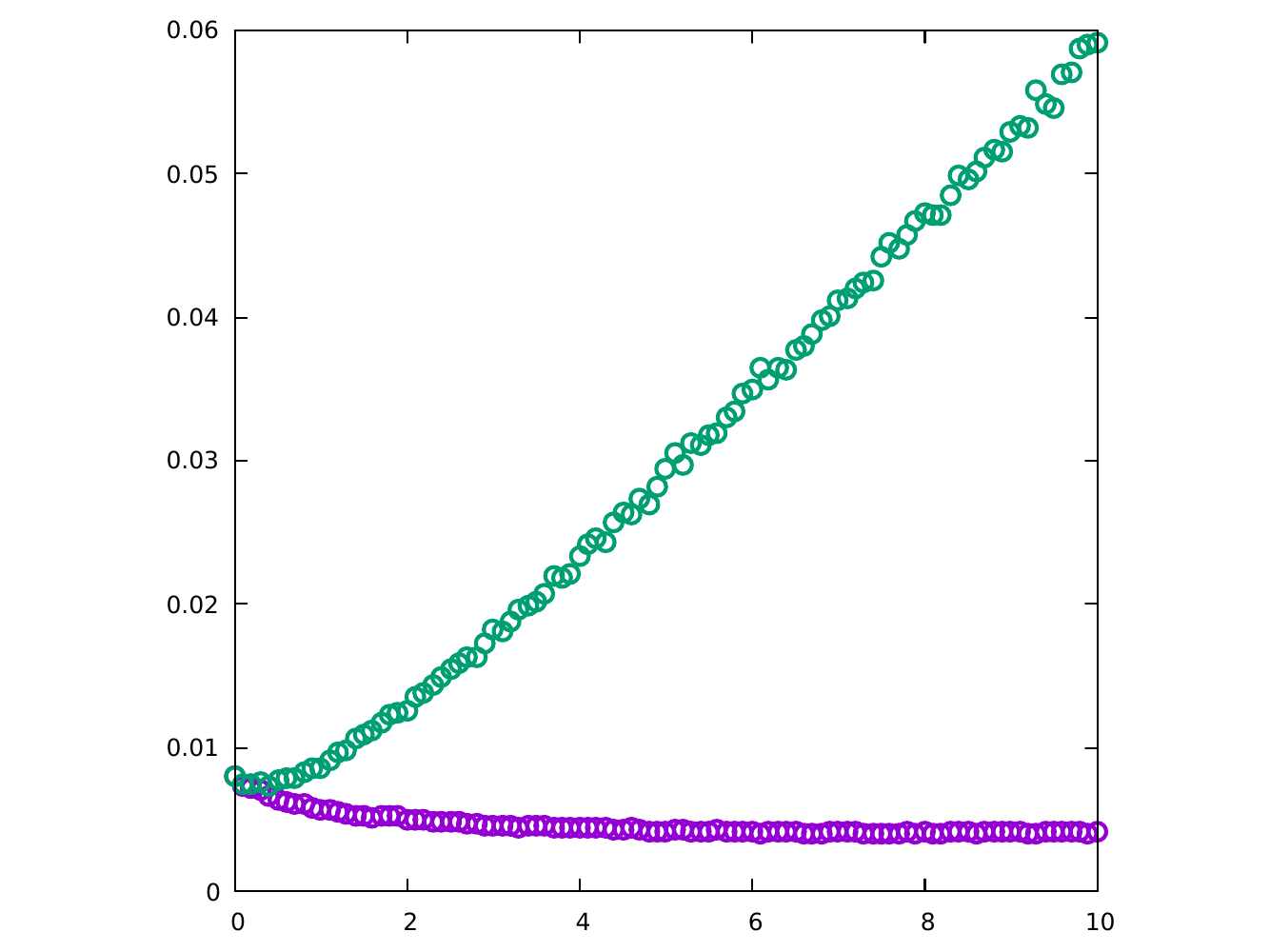}
    \caption{Plot of $\sigma_W^2$ versus $\Delta\Gamma$, for various $\phi_0$ : purple and green points are for $\phi_0=0,0.78$ respectively. For non-zero $\Delta\Gamma$, $k_\phi=50$. The isotropic fluctuation in work, however, is independent of the value of $k_\phi$.} 
\label{flucversusdelta1}    
\end{figure}

\begin{figure}[htp]
    \centering
    \includegraphics[width=8cm]{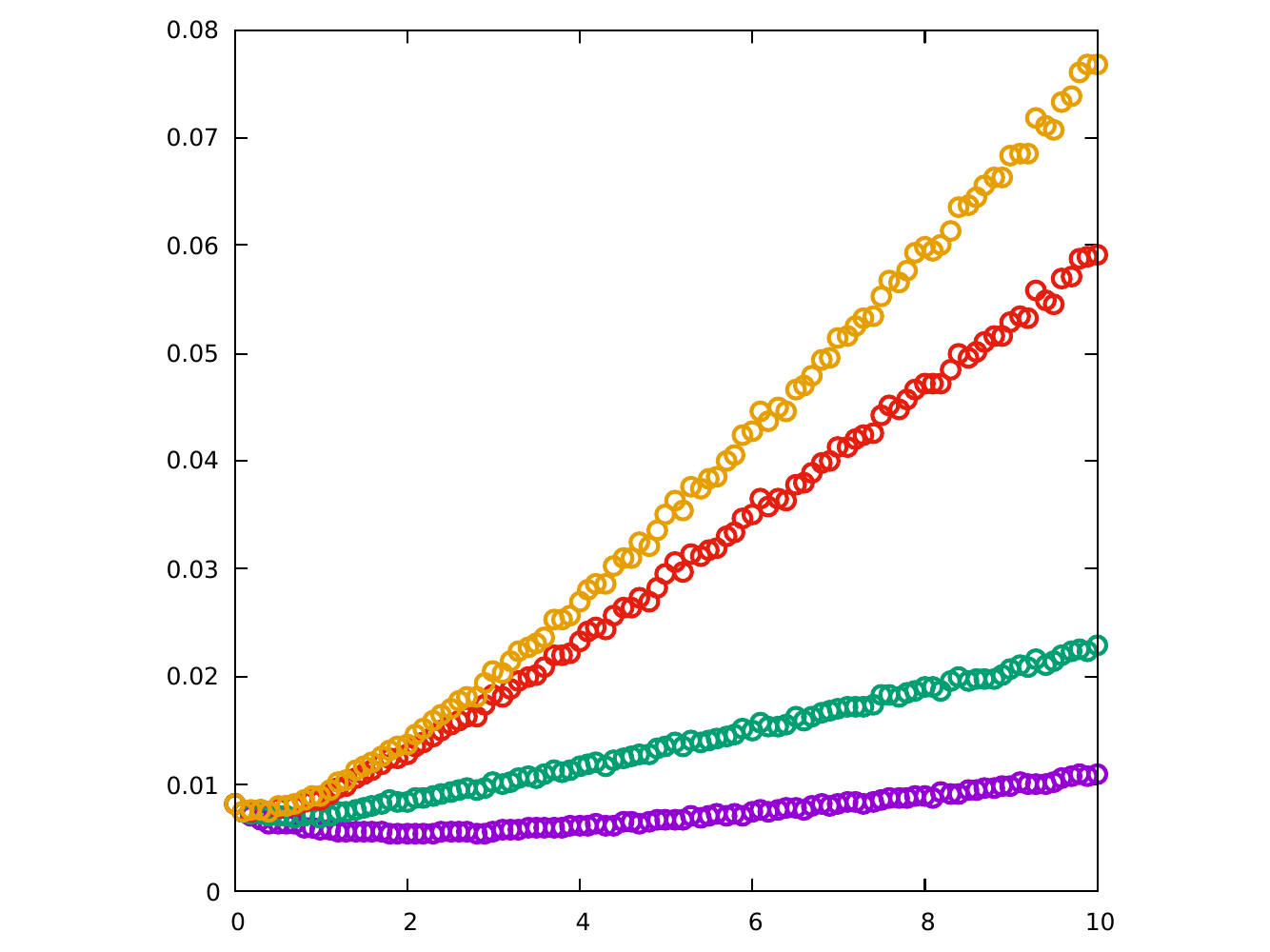}
    \caption{Plot of $\sigma_W^2$ versus $\Delta\Gamma$, for various non-zero $k_\phi$ : purple, green, red and yellow points are for $k_\phi=1,10,50,100$ respectively. Note that a minimum value of variance emerges (at smaller $\Delta\Gamma$), as $k_\phi$ in increased. For non-zero $\Delta\Gamma$, $\phi_0=0.78$. The isotropic fluctuation in work, however, is independent of the value of $\phi_0$.} 
\label{flucversusdelta2}    
\end{figure}

The variance of (or, fluctuation in) work is defined as, $\sigma_W^2\equiv \langle W^2\rangle - \langle W \rangle^2$, where $\langle\dots\rangle$ denotes a stochastic quantity averaged over several realizations (here, several successive engine cycles). In general, the extracted work is a fluctuating quantity, that solely depends on the stochastic trajectory availed by the working substance (see Eq.[\ref{jar}]). The variance gives the deviation of the stochastic work extracted after every cycle, from the average work that can be obtained by successively running these cycles. This is indeed an important quantity that scrutinizes whether each cycle is able to reproduce the value of average work, or how far is the work extracted in a cycle from the average value. For a stochastic heat engine, large fluctuations will cause different values of work to be extracted in different cycles, only a few being close to the average value. This is detrimental to the performance and output characteristics of such a machine because there will exist certain cycles which will yield a desirable work output as compared to others, thus causing a preferential hierarchy among various cycles. Hence, our aim will be to minimize this necessary evil with respect to the shape of the working substance, which will eventually render all the successive cycles to be equivalent. Fig.[\ref{flucversusdelta1}] shows the variation of this fluctuation with $\Delta\Gamma$, for various values of the mean orientation. For $\phi_0=0$, the variance saturates to an almost constant value, after slightly decreasing from the value at $\Delta\Gamma=0$. This shows that an ellipsoid with no orientational bias shows a lower fluctuation as compared to a sphere, when chosen as a working substance. The absence of dynamical coupling (on an average) is what causes this lower value of variance. However, at the MCR at $\phi_0=0$, the variance shows a considerable linear increase with $\Delta\Gamma$, after the isotropic value --- thus, maximal coupling of DoF can be seen as an inhibitory factor to the reproducibility of average work, whereas, it was responsible for maximizing the average extracted work. In such a situation, the value of $\Delta\Gamma$ can be tuned to be low (by changing the aspect ratio of the ellipsoid), such that the fluctuation remains at lower values. Next, the same variation needs to discussed with the other torque parameter, namely its strength $k_\phi$, as shown in Fig.[\ref{flucversusdelta2}]. After a brief non-linearity at smaller $\Delta\Gamma$, the variation becomes linear --- the slope of which increases with an increase in $k_\phi$, that is, the fluctuation in work increases more rapidly with the mobility difference of the ellipsoid, as the torque strength is increased. This is expected, as we have set $\phi_0$ at the MCR. As $k_\phi$ is increased, the thermal fluctuations in $\phi(t)$ are suppressed, leading to its distribution to be sharply crowded near $\phi_0=0.78$. This will cause a maximal coupling of the DoF (on an average), which is detrimental to the reproducibility of average work (as discussed earlier). The large variance in work coming from large $k_\phi$ can be counteracted by setting $\Delta\Gamma$ at the value where the variance shows a \enquote{shallow} minimum in the range $\Delta\Gamma\in [0:1]$ (see Fig.[\ref{flucversusdelta2}]). Hence, to optimize the performance of the engine (that is, to obtain maximum isotropic average work along with its better reproducibility after every cycle), $\phi_0=0.78$ can be complemented with large $k_\phi$ and small $\Delta\Gamma$. But then, a question arises --- how large a value of $k_\phi$ must be chosen? This will be decided in the next sub-section.

\subsubsection{Quasi-static, anisotropic efficiency $(\eta_\text{a})$}

The ratio of the absolute value of average extracted work and the average input heat (when the ellipsoid undergoes successive Stirling cycles) will give an estimate of how efficiently the anisotropic engine has converted the heat absorbed from the hot bath into useful thermodynamic work.  As discussed earlier, the efficiencies of both the isotropic as well as the anisotropic engine will lie below the Carnot bound, as dictated by the second law of thermodynamics. But then, it is also important to compare the efficiencies of the isotropic and anisotropic cases and how the latter can be improved. This sub-section aims to study the variation of anisotropic efficiency with the torque-strength $(k_\phi)$ and mobility difference $(\Delta\Gamma)$, also keeping in mind the need for optimizing the fluctuation in work.

\begin{figure}[htp]
    \centering
    \includegraphics[width=8cm]{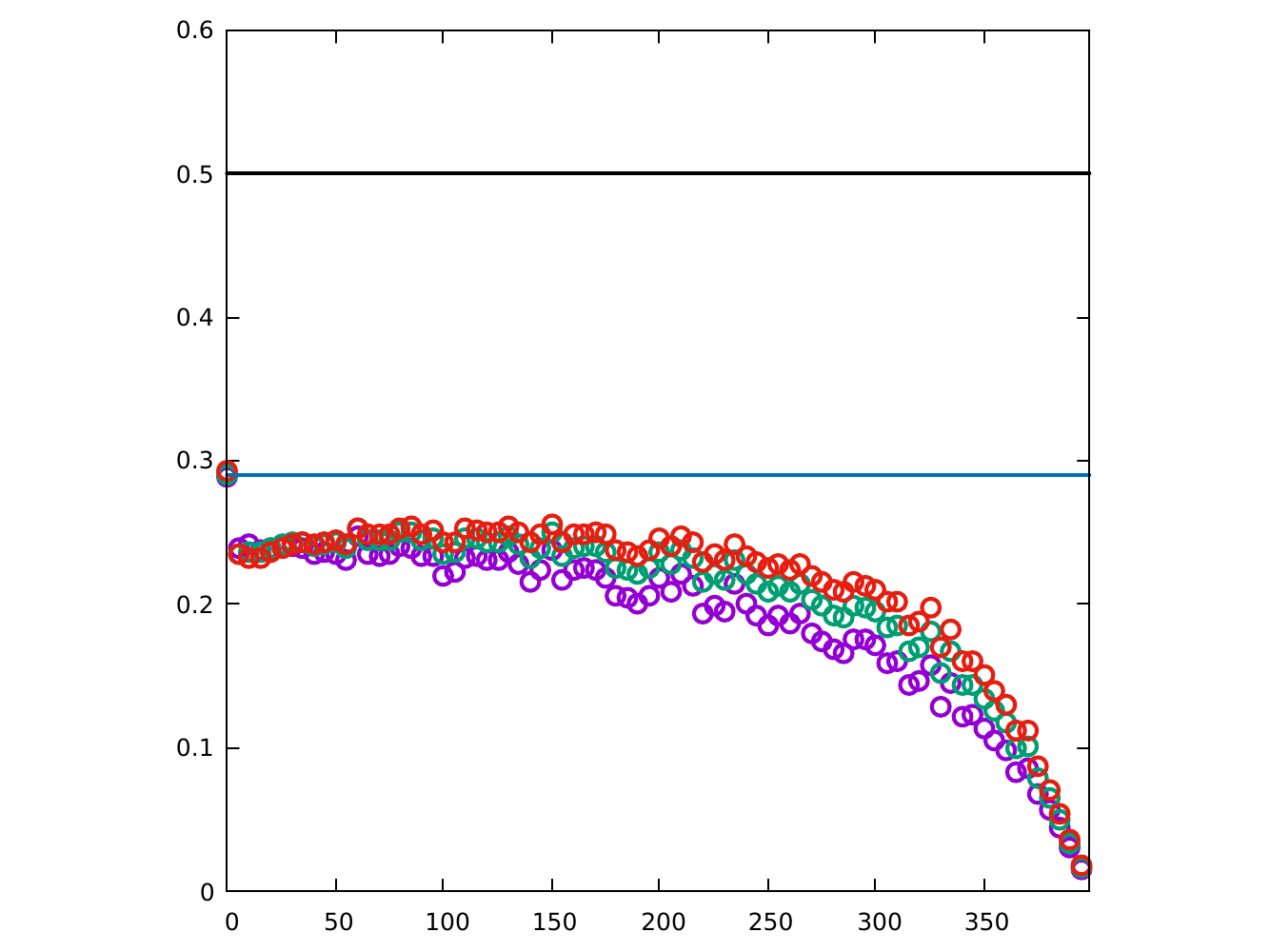}
    \caption{Plot of $\eta_\text{a}$ versus $k_\phi$ (at the MCR), for various non-zero $\Delta\Gamma$ : purple, green and red points are for $\Delta\Gamma=1,4,9$ respectively. The black and blue solid lines respectively denote the Carnot bound and isotropic benchmark.} 
\label{etaversuskphi}    
\end{figure}

In Fig.[\ref{etaversuskphi}], note that the anisotropic heat engine can never be more efficient than the isotropic one. We also see that $\eta_\text{a}$ merges with the isotropic value at the zero-torque limit, irrespective or $\Delta\Gamma$ (for reasons discussed earlier). As $k_\phi$ increases to large values, the efficiency decreases from a nearly constant value (which is more evident for a larger $\Delta\Gamma$) and gradually drops to zero for all the cases --- this was expected due to a large increase in the absorbed heat as compared to the extracted work (see Fig.[\ref{heatversuskphi}]). This can be seen as a \emph{heat wastage}, where a large amount of heat being absorbed (due to the internal energy associated with the torque) from the hot bath by the ellipsoid is not being recovered as useful work, but dissipated as unused energy. Roughly speaking, the range $k_\phi\in[50:150]$ is suitable to obtain an efficient engine in the anisotropic regime, complemented by a better reproducibility of the output.

\begin{figure}[htp]
    \centering
    \includegraphics[width=8cm]{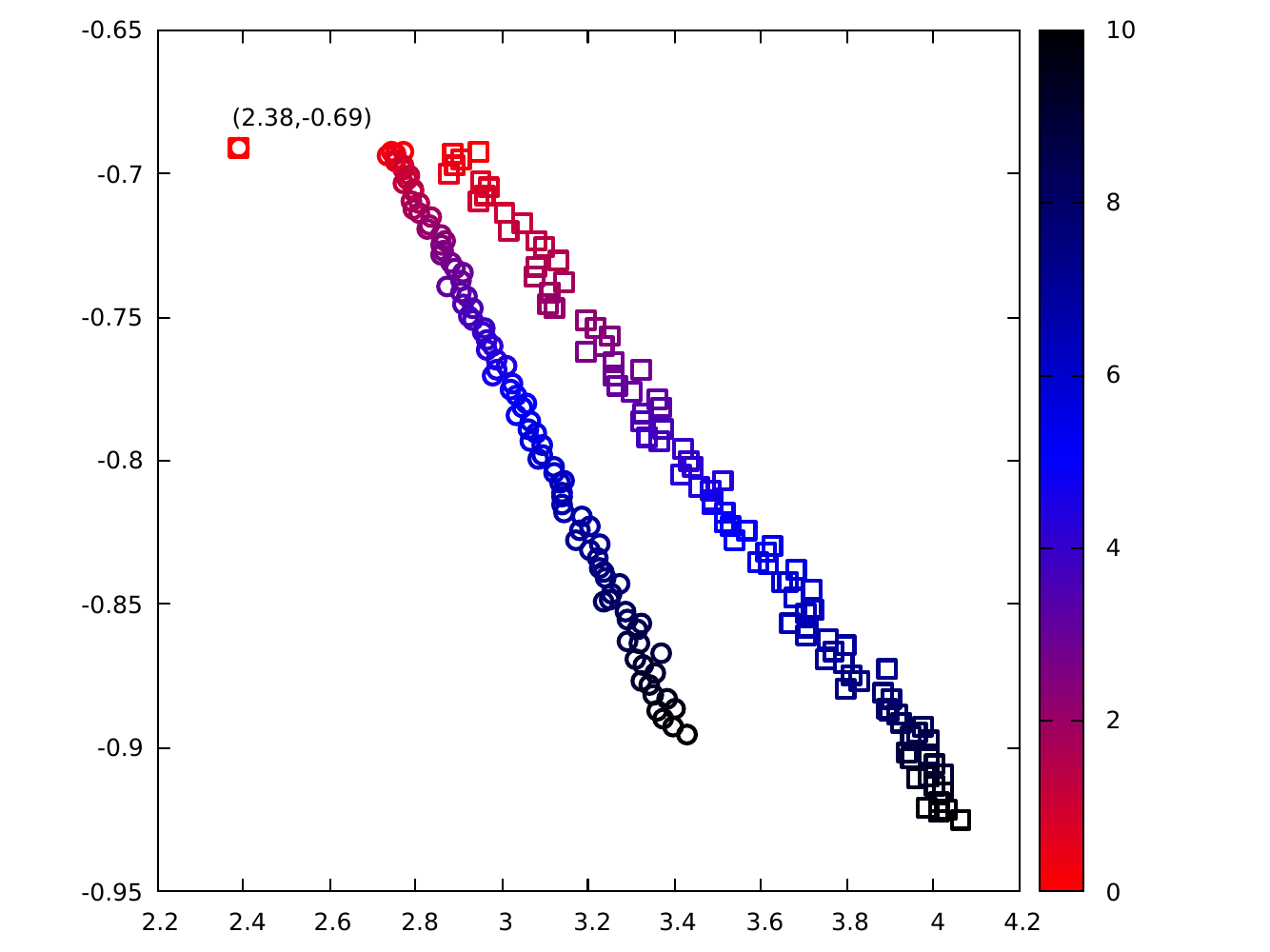}
    \caption{Plot of $\langle Q_1 \rangle$ along X-axis, $\langle W \rangle$ along Y-axis, and $\Delta\Gamma$ as color, for various $k_\phi$ : the circles and squares are for $k_\phi=1,10$ respectively. The isotropic values are labeled on the plot.} 
\label{simultaneous}    
\end{figure}

The same variation can be studied in the work versus heat plot. As shown in Fig.[\ref{simultaneous}], the extracted work decreases to larger negative values at the expense of an increase in the absorbed heat, w.r.t. $\Delta\Gamma$. This can be characterized by the linear region, developed after the isotropic limit. Conversely, the absolute value of the slope of this variation will give the efficiency --- we see that the slope slightly decreases as $k_\phi$ is increased.

\begin{figure}[htp]
    \centering
    \includegraphics[width=8cm]{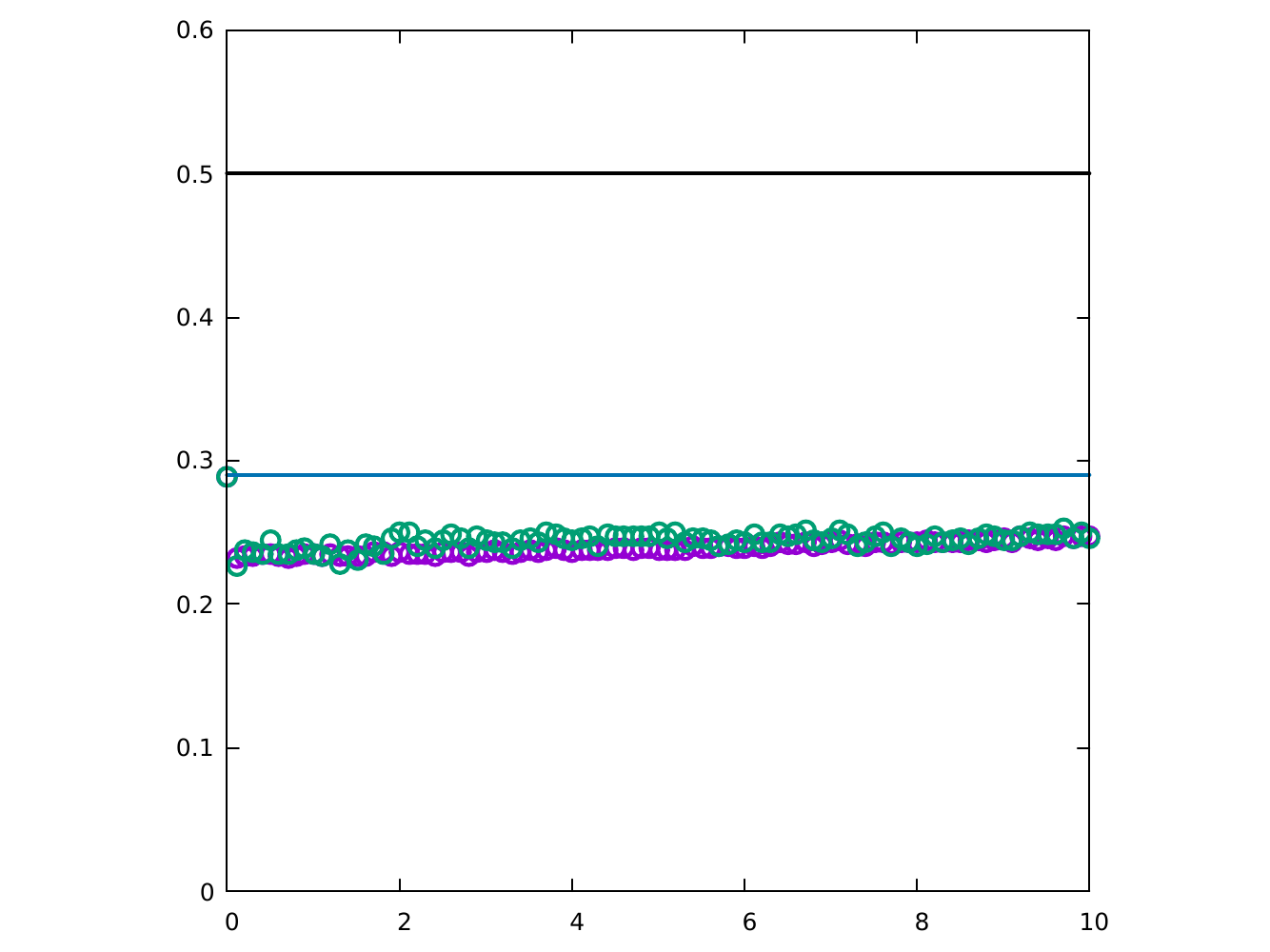}
    \caption{Plot of $\eta_\text{a}$ versus $\Delta\Gamma$, for various $\phi_0$ : purple and green points are for $\phi_0=0,0.78$ respectively. The black and blue solid lines respectively denote the Carnot bound and isotropic benchmark.} 
\label{etaversusdelta1}    
\end{figure}

\begin{figure}[htp]
    \centering
    \includegraphics[width=8cm]{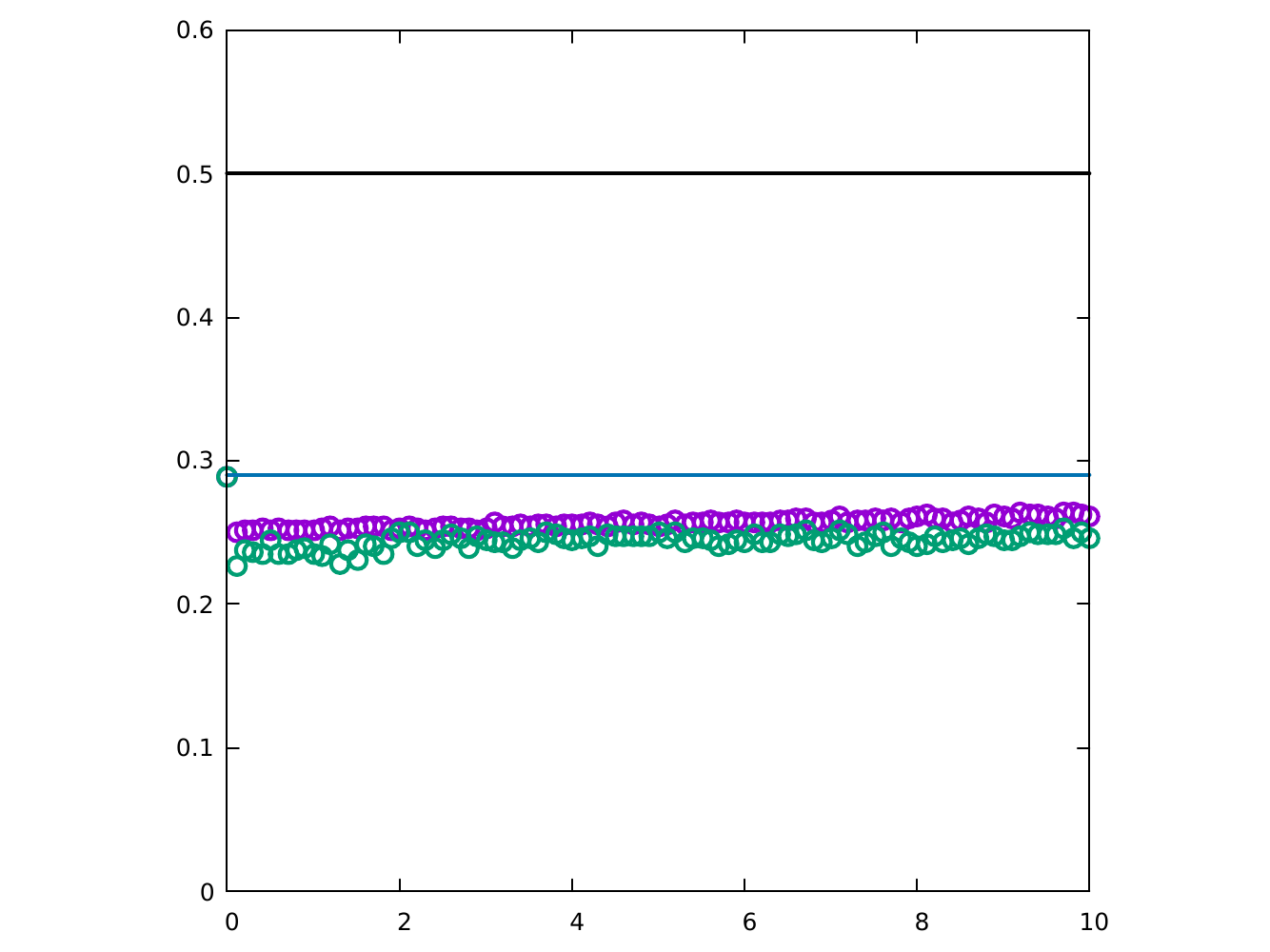}
    \caption{Plot of $\eta_\text{a}$ versus $\Delta\Gamma$, for various non-zero $k_\phi$ (at the MCR) : purple and green points are for $k_\phi=1,50$ respectively. The black and blue solid lines respectively denote the Carnot bound and isotropic benchmark.} 
\label{etaversusdelta2}    
\end{figure}

\begin{figure}[htp]
    \centering
    \includegraphics[width=9cm]{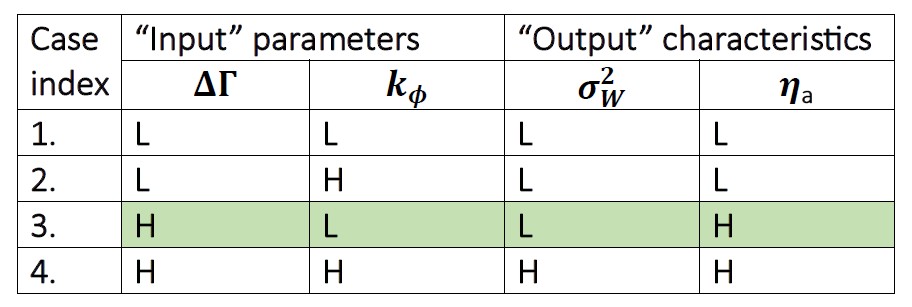}
    \caption{List of all the performance criteria, the primary factors affecting them, and the various inter-relationships (for $\phi_0=0.78$). The letters \enquote{L} and \enquote{H} are used to denote low and high values respectively (as per the system discussed, a high efficiency indicates values close to the isotropic limit). The third case is best suited for an overall optimized performance of the anisotropic engine.} 
\label{perfsummary}    
\end{figure}

Next comes the variation of $\eta_\text{a}$ with $\Delta\Gamma$. Fig.[\ref{etaversusdelta1}] shows that the isotropic efficiency is always greater than the anisotropic ones, where the latter shows a slight linear increase with $\Delta\Gamma$, coming closer to the former. But, the fluctuation in work is minimized at small $\Delta\Gamma$ --- so, a compromise between efficiency and reproducibility of output will be encountered for such an engine. Also, the efficiency is likely to be independent of the orientation of the ellipsoid, as evident from the merging of the various values of $\phi_0$. The same variation can now be studied for various torque-strengths, as shown in Fig.[\ref{etaversusdelta2}]. Beyond the isotropic regime, the lower $k_\phi$ starts at a higher value of $\eta_\text{a}$ and resides close to the isotropic limit, for reasons discussed earlier. Hence, if one wishes to design an engine in the higher-$\Delta\Gamma$ regime, the fluctuation in work output can be considerably reduced (along with the efficiency being close to the isotropic maximum) by keeping the restoring torque-strength lower (see Fig.[\ref{flucversusdelta2}]). A summary of all the performance criteria has been provided in Fig.[\ref{perfsummary}].

This concludes the section on the numerical results of the complete anisotropic system. In the next section, we will propose an approach to explicitly calculate the average extracted work by introducing physically-motivated approximations that will linearize the dynamical equations.

\section{Slightly anisotropic case --- Analytical results}

\subsection{Approximations (large $k_\phi$, small $\phi_0$, and small $\Delta\Gamma$) and linearized equations of motion}

For the coupling between $x$ and $y$ to survive statistically, we would need $\langle\sin\phi\cos\phi\rangle=\int \sin\phi\cos\phi P(\phi) d\phi \neq 0$, where $P(\phi)$ is the steady-state distribution of $\phi$ and the integral is taken over all possible values of $\phi$. Clearly, the distribution must have a broken inversion symmetry w.r.t. the angle $\phi$, that is, $P(\phi)\neq P(-\phi)$. This can be achieved by employing an external torque, as introduced in Eq.[\ref{eomm}] --- the net torque ($\tau$) fluctuates in time, such that $\langle\sin\phi\cos\phi\rangle\neq0$. Besides this, the non-zero value of $\phi_0$ helps in breaking the inversion symmetry of the corresponding distribution about $\phi=0$. Here, we will illustrate that if the steady-state orientational distribution is asymmetric, it will affect the translation of the particle by coupling its DoF.

A special case of the orientation-dependent torque is that of a restoring feature, the linear version of which is given by the Hooke's law : $-k_\phi (\phi-\delta\phi)$, where $k_\phi$ is the \enquote{strength} of this torque, and $\phi_0\equiv\delta\phi$ is the small angle to which the orientation is restored at large times. For large $k_\phi$, the typical time-scale of a given system for its translation $(\sim\frac{1}{k})$ becomes much greater than the time-scale for its rotation $(\sim\frac{1}{k_\phi})$. This means that a large value of $k_\phi$ suppresses the angular fluctuations, leading to a faster relaxation of the orientational DoF to its mean value, as fast as an exponential order. This leads to the fact that the translational dynamics of the ellipsoid can be averaged over all possible values of $\phi$. Such a suppression has another obvious consequence --- the steady-state variance of $\phi$ (that is, the \enquote{width} of the distribution $P(\phi)$) becomes considerably small. Thus, only the first-order moment of $\phi$ has the dominant contribution, and the second and other higher-order moments are quite small in comparison to its first moment $(\delta\phi)$.

One can further simplify the dynamical equations by taking the limit $\Delta\Gamma\rightarrow 0$, but not exactly equal to zero \cite{mandal2024diffusion}. In this limit, $\gamma_{\perp}$ and $\gamma_{\parallel}$ are so close to each other that both can be replaced by their average, $\gamma=\frac{\gamma_{\perp}+\gamma_{\parallel}}{2}$. The limit is valid for particles having a shape which slightly deviates from a perfect spheroidal symmetry. This can be practically devised by tuning the \emph{aspect ratio} ($\alpha$) of the ellipsoidal particle, which is simply the ratio of the lengths of its long and short principal axes (see Fig.[\ref{ellipsoid}]). Aspect ratio can also be thought of as a measure of the shape anisotropy of the particle, where, $\alpha=1$ is the spherical limit. Using the explicit expressions for longitudinal and transverse friction coefficients \cite{berg1993random}, one can find the ratio of the mobility difference and mean mobility in terms of the aspect ratio ($\alpha>1$ for a prolate ellipsoid) as : $\frac{\Delta\Gamma}{\Gamma}=\frac{\ln2\alpha - 1.5}{3\ln2\alpha-0.5}$. To keep this ratio less than unity, the value of $\alpha$ can be suitably tuned for fabricating an ellipsoid of a definite shape.

Using these approximations, one can taken an average on both sides of Eq.[\ref{eom2}] w.r.t. $P(\phi)$ (that is, integrated over all possible values of $\phi$), to obtain the following set of linear differential equations governing the simplified, $\phi$-averaged, effective translational dynamics of the ellipsoid, with all the other qualitative features remaining intact:

\begin{eqnarray}
\nonumber
\dot{x}=-\frac{k(t)}{\gamma} (x+\epsilon y)+\sqrt{\frac{2T}{\gamma}}\zeta_x(t)\\
\dot{y}=-\frac{k(t)}{\gamma} (y+\epsilon x)+\sqrt{\frac{2T}{\gamma}}\zeta_y(t)
\label{eomapprox}
\end{eqnarray}

where, $\epsilon=\left(\frac{\Delta\Gamma}{\Gamma}\right)\delta\phi$ is the dissipative coupling parameter $(<<1)$, and $\langle\zeta_i(t)\zeta_j(t')\rangle=\delta_{ij}\delta(t-t')$ with $\langle\zeta_i(t)\rangle=0$. Here, $\Gamma=\frac{\Gamma_{\perp}+\Gamma_{\parallel}}{2}$ is the mean mobility, which is obviously greater than $\Delta\Gamma$, subjected to the above limit. Since the parameter $\epsilon$ is the product of two quantities whose respective magnitudes are less than unity, it must be set in such a way that the dynamic coupling of the DoF remains undisturbed.

It is evident that the coupling between $x$ and $y$ survives in Eq.[\ref{eomapprox}] through the geometric anisotropy $(\Delta\Gamma)$ of the Brownian ellipsoid and the non-zero, average orientation $(\delta\phi)$, both being the intrinsic features of the particle itself. As the equations are linear, the distributions of $x(t)$ and $y(t)$ will be Gaussian (due to the involved noises). We emphasize here that the approximation regarding the small fluctuations in $\phi$ is not mandatory to obtain useful thermodynamic work --- this simplification reduces the number of system parameters in the analysis. In the following few sub-sections, we will show that a Brownian ellipsoid can act as a working substance even in the limit of small-coupling, while discussing the method to obtain an expression for the average extracted work. For simplicity, we will consider Eq.[\ref{eomapprox}] as the equations of translational motion of the center of mass of the ellipsoid.

\subsection{ Position moments}

Eq.[\ref{eomapprox}] can be equivalently expressed in terms of matrices as:

\begin{eqnarray}
    \boldsymbol{\dot{R}}=-\frac{k(t)}{\gamma}{\bar{\bar{C}}}\boldsymbol{R}+\sqrt{\frac{2T}{\gamma}}\boldsymbol{\zeta}(t)
    \label{matrixeqn}
\end{eqnarray}

where, $\boldsymbol{R}=\{x,y\}$ is a vector containing the coordinates of center of mass of the ellipsoid in the lab frame, and $\boldsymbol{\zeta}(t)=\{\zeta_x(t),\zeta_y(t)\}$ is a corresponding vector containing the translational noises. The $2\times2$ matrix, $\bar{\bar{C}}=\begin{bmatrix} 1 & \epsilon \\ \epsilon & 1 \end{bmatrix}$, is the so-called \enquote{coupling tensor}.

Calculating the position moments, while they are dynamically coupled, is non-trivial. This becomes easier if one diagonalizes the coupling tensor to shift to a system of transformed co-ordinates, where the DoF are decoupled. Eventually, an inverse transformation can lead to the moments we need for the calculation of average extracted work. The matrix, $S=\frac{1}{\sqrt{2}}\begin{bmatrix} -1 & 1 \\ 1 & 1 \end{bmatrix}$, will diagonalize the coupling tensor. On multiplying both sides of Eq.[\ref{matrixeqn}] with $S^{-1}$ from left and using the orthogonality relation, $SS^{-1}=S^{-1}S=\mathcal{I}$ (with $\mathcal{I}$ being the identity matrix), we can obtain a set of two decoupled differential equations in the transformed coordinate system:

\begin{eqnarray}
    \nonumber
    \dot{\tilde{x}}=-\frac{k(t)}{\gamma} (1-\epsilon) \tilde{x}+\zeta_{\tilde{x}}(t)\\
    \dot{\tilde{y}}=-\frac{k(t)}{\gamma} (1+\epsilon) \tilde{y}+\zeta_{\tilde{y}}(t)
    \label{decoupled}
\end{eqnarray}

where, the transformed quantities (denoted by an overhead tilde) and the original ones bear the following relations:

\begin{eqnarray}
    \nonumber
    \tilde{x}&=&\frac{1}{\sqrt{2}}(-x+y)\\
    \nonumber
    \tilde{y}&=&\frac{1}{\sqrt{2}}(x+y)\\
    \nonumber
    \zeta_{\tilde{x}}(t)&=&\sqrt{\frac{T}{\gamma}}[-\zeta_x(t)+\zeta_y(t)]\\
    \zeta_{\tilde{y}}(t)&=&\sqrt{\frac{T}{\gamma}}[\zeta_x(t)+\zeta_y(t)]
\end{eqnarray}

The respective FDRs obeyed by the transformed noises can be evaluated using the above relations. Eq.[\ref{decoupled}] are a set of linear differential equations, the general solutions to which can be formally written in terms of an integrating factor. These can be used to obtain the moments of position in the transformed coordinate system. Having obtained the same, one can simply plug them in the following inverse transformation relations, to obtain the position moments (for slight anisotropy) in the original system of coordinates: $\langle x^2 \rangle = \langle y^2 \rangle = \frac{1}{2}[\langle \tilde{x}^2 \rangle + \langle \tilde{y}^2 \rangle]$, $\langle xy \rangle = \frac{1}{2}[\langle \tilde{y}^2 \rangle - \langle \tilde{x}^2 \rangle]$, which can now be given as:

\begin{eqnarray}
    \nonumber
    \langle x^2 \rangle = \langle y^2 \rangle &=& \frac{T}{k(t)(1-\epsilon^2)}\\
    \langle xy \rangle &=& -\frac{T}{k(t)}\left(\frac{\epsilon}{1-\epsilon^2}\right)
    \label{momentsaniso}
\end{eqnarray}

Eq.[\ref{momentsaniso}] shows that the position moments in the isotropic case (see Eq.[\ref{moments}]) have been modulated by a shape-dependent factor in the small-coupling regime, where $\epsilon=0$ reproduces the isotropic results. Also, for the complete anisotropic system, the saturation of cross-correlation to negative values for large $k_\phi$ (as shown in Fig.[\ref{corr2}]) can now be reconciled with the result obtained in the slightly anisotropic case --- where, $\epsilon=0$ yields the isotropic limit, where there exists no dynamical correlation between the DoF. However, it must be kept in mind that the small-coupling limit can not capture all the nuances of the complete isotropic system, and the validity of the associated approximations must be tested.

\subsection{Average extracted work --- validity of approximations}

\begin{figure}[htp]
    \centering
    \includegraphics[width=9cm]{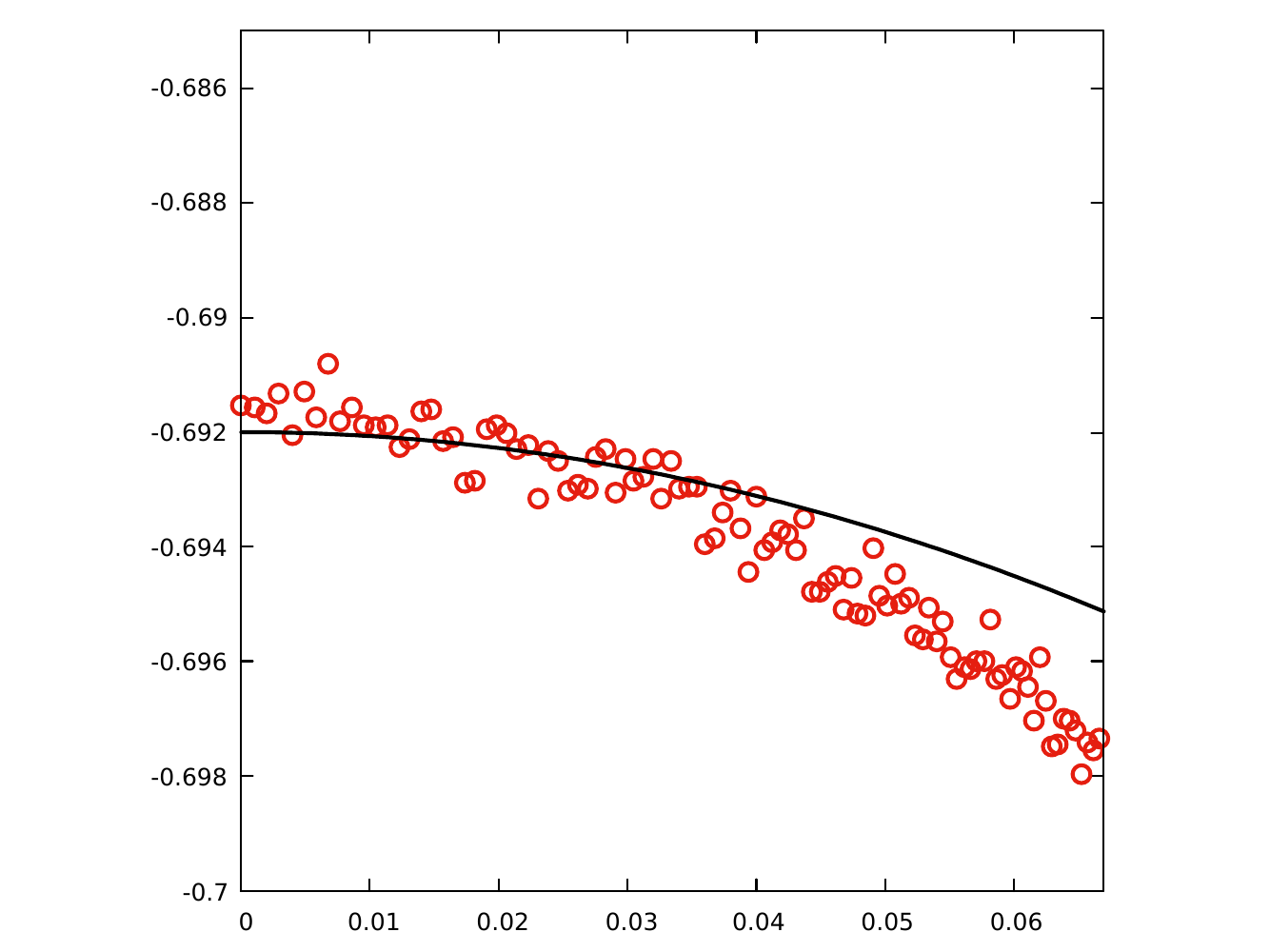}
    \caption{Plot of the average extracted work versus $\epsilon$, for $k_\phi=50$ and $\phi_0=0.1$. The values of $\epsilon$ are generated by keeping $0\leq\Delta\Gamma\leq1$. The red points are obtained from the numerical simulation of the complete anisotropic system, and the black solid line is obtained from Eq.[\ref{slightwork}].} 
\label{agreement}    
\end{figure}

Plugging the position moments (given in Eq.[\ref{momentsaniso}]) in Eq.[\ref{workdef}], we can obtain the average work extracted in the slightly anisotropic limit $(\epsilon<<1)$, w.r.t. the protocol introduced earlier:

\begin{eqnarray}
    \langle W \rangle_\text{\it{sa}} = \frac{\langle W \rangle_\text{\it{i}}}{1-\epsilon^2}
    \label{slightwork}
\end{eqnarray}

where, the labels {\it{sa}} and {\it{i}} stand for \enquote{slightly anisotropic} and \enquote{isotropic} respectively. Clearly, $\epsilon=0$ yields the value of isotropic work, $\langle W \rangle_\text{i}$, as defined in Eq.[\ref{isowork}], and the work shows a decrease from this value as $\epsilon$ increases from zero. This non-linear region was also manifested in Fig.[\ref{versusdelta1},\ref{versusdelta2}] at smaller values of $\Delta\Gamma$. The fractional change (or, relative error) in work, as one goes from the isotropic case to the slightly anisotropic regime, is equal to $\frac{\epsilon^2}{1-\epsilon^2}$, which clearly diminishes for $\epsilon\to0$. Fig.[\ref{agreement}] shows that the numerical result for the complete anisotropic system (with large $k_\phi$, small $\phi_0$, and small $\Delta\Gamma$) agrees well with the analytical result for the slightly anisotropic case only for values of $\epsilon$ much less than unity. As the value of $\epsilon$ is increased, this agreement becomes poorer, as expected. Nonetheless, both the results agree at the isotropic benchmark.

\section{Concluding remarks and further prospects}

To conclude, we will summarize the results here. We have considered a Brownian ellipsoid in 2D, trapped in a harmonic confinement that produces restoring force as well as torque. A dissipative coupling between the space coordinates arises due to the difference between the longitudinal and transverse mobilities and a finite mean orientation of the ellipsoid, ensured by the restoring torque. The stiffness of the restoring force is changed time-periodically and quasi-statically to imitate the strokes of a piston. This piston-like protocol drives the system through the four strokes of a Stirling cycle, where the ambient fluid is alternately connected to two thermal reservoirs at different temperatures. In the midst of these ingredients of a stochastic heat engine, the single ellipsoid acts as a working substance, taking heat from the hot bath and partially converting it into useful thermodynamic work. After calculating the relevant thermodynamic quantities in the isotropic limit, we have numerically studied the full anisotropic system. Both the extracted work as well as the input heat have been shown to explicitly depend on the geometry of the ellipsoid and the parameters that control its orientation. The performance of the engine has been quantified by computing the fluctuation in work and quasi-static efficiency, the latter being bounded by both the Carnot limit as well as the isotropic benchmark. Different ways have been prescribed to yield higher efficiency at lower fluctuation. Finally, we have dealt with the slightly anisotropic case, where the dynamical equations have been linearized using physically-motivated approximations. The average extracted work has been calculated in this small-coupling limit, which shows an agreement with the numerical results of the fully anisotropic system, when the parameters are kept within the stipulated range. From this, it can be inferred that a passive, anisotropic particle can generate work by rectifying its ambient thermal fluctuations, and the resultant system can work as a stochastic engine. More significantly, the current study shows that the energetics of the particle exhibit explicit dependence on the geometry of the working substance and the dependence occurs primarily due to the dissipation, which directly involves the geometry of the particle.

The trapped Brownian ellipsoid can be considered as the simplest candidate of the entire class of bi-axial particles. One can then further explore the overall relationship between the stochastic thermodynamic quantities and the inherent geometry of the system involved in the dynamics, for more generic cases. Finally, as it happens inherently in the living world, one can address the issue that if the geometry of the system fluctuates over space and time, how the stochastic energetics of the system will vary depending on these fluctuations. This is an open question in the contemporary research in this field, which can now be studied with the current work as the initiating juncture.

\section*{Acknowledgements}

A.S. acknowledges the Core Research Grant (CRG/2019/001492) from DST, Government of India. A.S. also acknowledges the funding (\emph{Investissements d’Avenir}, ANR-16-IDEX-0008) by CY Initiative of Excellence. The present work was partially developed during his stay at the CY Advanced Studies, whose support is acknowledged. \\

{\section*{References}}
\bibliographystyle{unsrt}
\bibliography{sample.bib}

\end{document}